\documentclass[journal]{IEEEtran}

\usepackage{amsmath,amssymb,amsthm,bm}
\usepackage{algorithm}
\usepackage{algpseudocode}
\usepackage{array}
\usepackage{xcolor}
\usepackage{cite}
\usepackage{mathrsfs}
\usepackage{graphicx}
\usepackage{booktabs}
\usepackage{mathtools}
\usepackage{microtype}
\usepackage{url}

\newtheorem{theorem}{Theorem}
\newtheorem{lemma}{Lemma}
\newtheorem{proposition}{Proposition}
\newtheorem{corollary}{Corollary}
\newtheorem{remark}{Remark}
\newtheorem{assumption}{Assumption}

\newcommand{\R}{\mathbb{R}}
\newcommand{\C}{\mathbb{C}}

\providecommand{\revisioncolor}{}
\providecommand{\revstart}{\begingroup\revisioncolor}
\providecommand{\revstop}{\par\endgroup}
\providecommand{\rev}[1]{{\revisioncolor #1}}
\begin{document}

\title{Energy-Harvesting-Aware Cell-Free LEO Satellite Networks: A Lyapunov Framework for Sustainable User-Centric Beamforming and Clustering}

\author{Wali Ullah Khan \thanks{Wali Ullah Khan is with the School of Computing, Gachon University, Seongnam-si 13120, Republic of Korea (email: waliullahkhan30@gmail.com).

}}

\maketitle

\begin{abstract}
\revstart
We study battery-aware downlink control in user-centric cell-free low Earth orbit satellite networks with intermittent solar harvesting and finite onboard batteries. A perturbed Lyapunov controller coordinates clustering and hybrid beamforming using measured battery states and causal harvest forecasts. The per-slot solver combines group-sparse association, rate bounds, a quadratic transform, and successive convex approximation. We establish robust battery invariance for feasible implemented actions and give a utility bound conditional on a comparison policy and the per-slot implementation gap. Multi-orbit control-layer simulations show that queue-aware scheduling preserves a larger battery reserve in the stressed scenario at the expense of effective energy efficiency. A separate fixed-cluster matrix snapshot illustrates recovered hybrid beam patterns and the per-slot objective trace.
\revstop
\end{abstract}


\section{Introduction}


Low Earth orbit (LEO) constellations are becoming a central component of emerging non-terrestrial networks because they provide lower propagation delay and lower free-space path loss than geostationary satellite systems~\cite{11432996,10097680}. Current multi-beam deployments, however, remain predominantly satellite-centric: each user equipment (UE) is associated with a single spot beam, while mobility is handled through repeated beam and satellite handovers~\cite{11295961,11435637,11003898}. The rapid orbital motion of LEO satellites further complicates the acquisition and maintenance of instantaneous channel state information (CSI) and increases handover and signaling overhead \cite{11366110,11526722,11244159}.

User-centric cell-free NTN architectures offer a promising alternative by allowing each UE to be jointly served by a geometry-dependent cluster of satellites \cite{10966047,10787138}. In such architectures, clustering and precoding decisions can be updated using orbital ephemeris and UE location, relying mainly on slowly varying statistical CSI rather than rapidly changing small-scale fading realizations~\cite{11037747,11145289}. This architecture can improve macro-diversity, mobility robustness, and service continuity. At the same time, cooperative transmission couples the energy expenditure of multiple satellites, because clustering, beamforming, data forwarding, and RF-chain activation jointly determine the electrical load drawn from their onboard power systems \cite{10396846}.

Most existing cell-free NTN resource-allocation frameworks impose a fixed per-satellite transmit-power budget. Although convenient analytically, such a constraint is not a physical model of satellite energy availability \cite{11592460}. A LEO satellite harvests solar energy only during illuminated portions of its orbit, experiences recurrent eclipse intervals, and supplies communication, thermal, attitude-control, telemetry, housekeeping, and inter-satellite-link loads from the same finite battery \cite{11478007,11586009,11045394}. A controller that repeatedly exploits a fixed RF power limit may progressively deplete the battery before or during eclipse, whereas an overly conservative static cap may leave available solar energy unused or cause avoidable battery overflow \cite{9492855}.

The resulting design problem is inherently temporal. A clustering or beamforming decision made in the current slot changes the battery state and therefore alters the feasible action set in subsequent slots. Consequently, enforcing an instantaneous power constraint alone is insufficient: the controller must account for the opportunity cost of consuming stored energy now rather than preserving it for future eclipse periods.

To address this coupling, we construct a perturbed Lyapunov queue directly from the measured physical battery state. The queue acts as a time-varying marginal value of onboard energy. When a satellite operates below its target reserve, the queue penalizes communication load and promotes energy conservation; when the battery is above the target, it encourages the use of otherwise spillable harvested energy. Hence, clustering and beamforming are selected not only according to current channel conditions and instantaneous power availability, but also according to the intertemporal value of stored energy.

This work therefore embeds user-centric clustering and hybrid beamforming within an explicit solar-harvesting and finite-battery model. The controller must exploit favorable communication geometry, preserve battery reserve across eclipse intervals, satisfy physical energy-causality constraints, and remain compatible with the nonconvex fractional structure of cell-free hybrid beamforming.

\subsection{Related Work}
\label{sec:related}


Extensive prior work has investigated beam management, resource allocation, handover control, and precoding for satellite-centric NTN architectures operating under the 3GPP multi-beam coverage model~\cite{3gpp38821,azari2022evolution}. This includes surveys of precoding techniques for high-throughput satellite systems~\cite{khammassi2023precoding} and stochastic-geometry-based performance analyses of LEO satellite links~\cite{talgat2021stochastic}. More recently, user-centric cell-free NTN designs have replaced fixed satellite--UE attachment with dynamically formed cooperative clusters. In particular, the cell-free massive non-terrestrial network architecture in~\cite{kim2025cellfree} uses geometry-driven statistical CSI to coordinate multiple satellites, thereby improving macro-diversity and mobility management while reducing the signaling burden associated with instantaneous-CSI acquisition.


Building on the cell-free NTN paradigm, a natural objective is to optimize network energy efficiency rather than treating it as a secondary outcome of rate maximization. This requires jointly designing satellite clustering and hybrid analog--digital beamforming subject to transmit-power, RF-chain, and quality-of-service (QoS) constraints. Such problems are typically nonconvex, fractional, and combinatorial. Relevant methodological tools include the quadratic transform for fractional programming in communication systems~\cite{shen2018fractional}, the Charnes--Cooper transformation for linear-fractional optimization~\cite{charnes1962programming}, reweighted-$\ell_1/\ell_2$ relaxations for sparsity-promoting association design~\cite{candes2008enhancing}, and Riemannian optimization on the complex-circle manifold for constant-modulus RF precoding~\cite{absil2009optimization,sato2022riemannian}. Recent work has continued in this direction; for example,~\cite{thu2026joint} jointly optimizes load balancing and transmit power to improve uplink energy efficiency in a hybrid satellite cell-free massive-MIMO system.

Across these formulations, however, the per-satellite transmit-power budget is generally treated as a fixed design parameter, independent of the satellite's instantaneous battery state, solar-harvesting condition, or proximity to an eclipse interval. Consequently, the optimization captures communication-side energy efficiency but not the temporal evolution of the onboard electrical power subsystem. Lyapunov drift-plus-penalty control provides a general framework for transforming long-term stochastic optimization problems into a sequence of online per-slot decisions~\cite{neely2010stochastic}. By balancing instantaneous utility against the drift of suitably defined queues, the framework \rev{can yield utility and queue trade-offs under suitable comparison and feasibility conditions.}

Lyapunov optimization has been widely applied to terrestrial energy-harvesting wireless networks, in which transmission and processing decisions draw energy from a stochastic source buffered by a finite battery. For example,~\cite{huang2013utility} develops utility-optimal online scheduling under long-term energy-causality constraints. These methods show that future energy availability can be accounted for without requiring noncausal knowledge of the harvesting process. To the best of our knowledge, this framework has not previously been integrated with the joint satellite-clustering and hybrid-beamforming problem in user-centric cell-free NTNs. The present setting is structurally distinct from terrestrial energy-harvesting systems because the harvested energy follows an orbital sunlight--eclipse pattern, while each communication action simultaneously affects multi-satellite cooperation, RF-chain usage, forwarding load, and future battery feasibility.


The satellite systems engineering literature has long studied power generation, battery storage, eclipse operation, and electrical power subsystem sizing~\cite{wertz1992space,khan2020comprehensive}. More recent works have investigated online and learning-based energy management specifically for LEO satellites. These include Q-learning-based transmit-power control for extending battery lifetime~\cite{tsuchida2020efficient}, online convex energy scheduling for reducing battery depth of discharge in satellite edge computing~\cite{li2022towards}, and a Lyapunov-model-predictive-control strategy that balances predicted solar harvesting against anticipated service demand for a single LEO satellite~\cite{wang2025green}.

These studies demonstrate the operational importance of battery-state-aware control. However, they consider a single satellite or a computation-oriented energy-management problem and do not address cooperative multi-satellite transmission in which clustering and beamforming decisions jointly determine the load drawn from several finite batteries. Conversely, existing cell-free NTN optimization studies coordinate multiple satellites but generally treat their transmit-power limits as static and independent of orbital energy dynamics. Our work bridges these two research directions by coupling physical battery states, sunlight--eclipse evolution, user-centric clustering, and hybrid beamforming through an online Lyapunov control framework. 

\begin{figure*}[!t]
    \centering
    \includegraphics[width=0.8\textwidth]{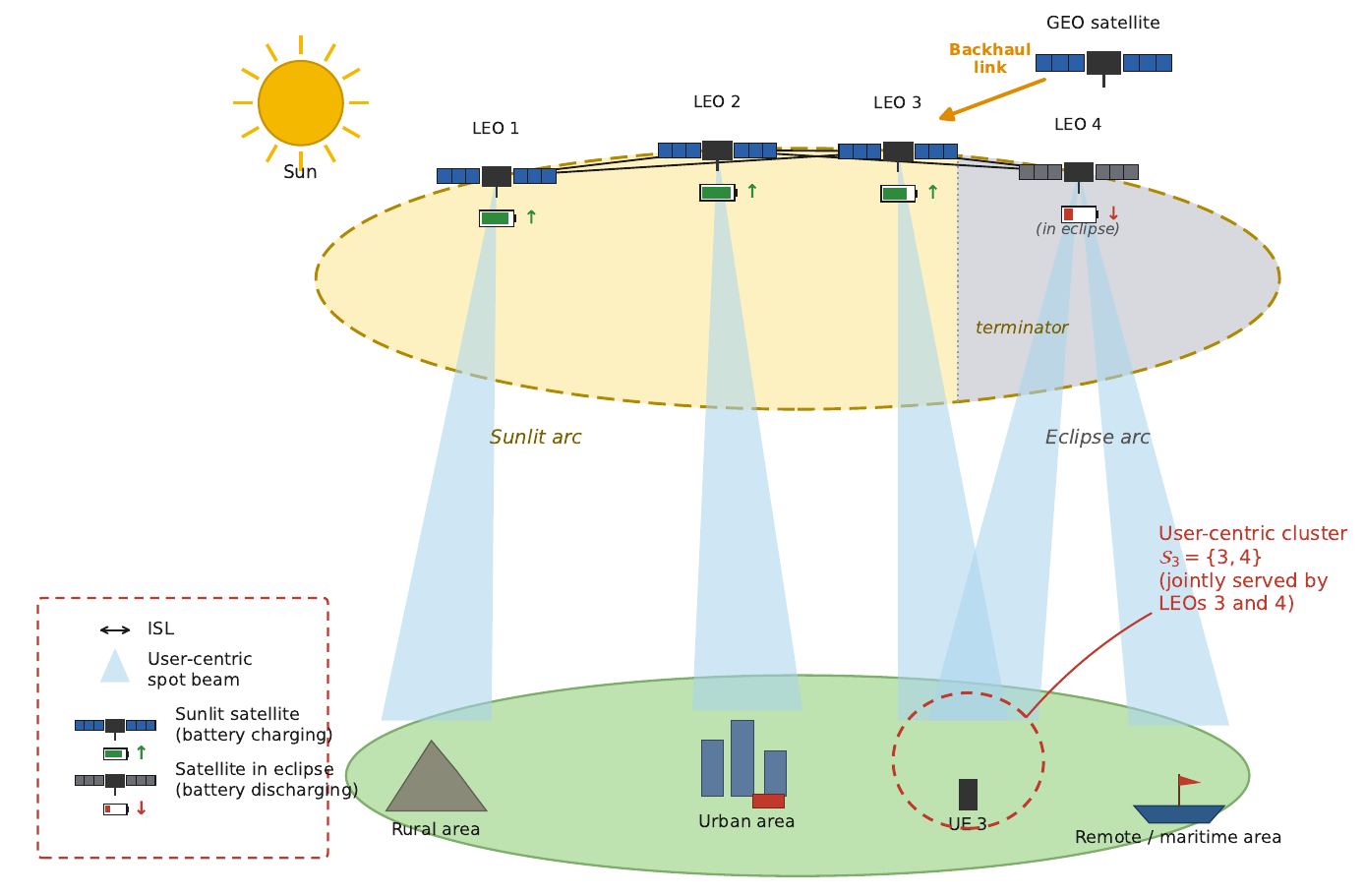}
    \caption{System model of the considered solar-powered user-centric
    cell-free LEO satellite network. LEO satellites cooperate through
    inter-satellite links and may connect to a GEO satellite through a
    backhaul link. Satellites operating along the sunlit portion of the
    orbit harvest solar energy and charge their onboard batteries, whereas
    satellites in eclipse draw energy from their batteries. Each ground
    user is served by a dynamically formed satellite cluster through
    overlapping spot beams. The highlighted example shows UE~3 jointly
    served by LEOs~3 and~4, corresponding to
    $\mathcal{S}_3=\{3,4\}$.}
    \label{fig:system_model}
\end{figure*}

\revstart
Recent satellite-assisted cell-free work studies dynamic clustering and robust beamforming under transmit-power constraints~\cite{10966047}. Lyapunov methods have also been applied to on-orbit computing and task offloading~\cite{yan2026orbit}. Here the queue is the signed deviation of the measured physical battery from a reserve target, and its weight enters the cooperative downlink decision. Nonlinear vibration harvesting is another branch of energy harvesting. Bistable piezoelectric structures can collect energy over a wider mechanical excitation band~\cite{arrieta2010plate}. The modeled source here is orbital photovoltaics, whose availability follows illumination and eclipse.
\revstop
\subsection{Main Contributions}
\revstart
\begin{itemize}
\item We incorporate the marginal value of stored energy into user-centric clustering and beamforming while accounting for platform and forwarding loads and bounded solar forecast errors.
\item We construct a per-slot solver with sign-dependent load bounds and separate the guarantees for the continuous subproblem, implemented battery safety, and long-run utility. We also distinguish operating-point sensitivity to energy budgets from first-order sensitivity to satellite position.
\item We evaluate the reserve--efficiency trade-off over multiple orbits and illustrate spatial selectivity and solver behavior in a controlled matrix-level snapshot.
\end{itemize}
\revstop
The remainder of this paper is organized as follows. Section~II presents the network, channel, onboard-power, solar-harvesting, and battery models, together with the main assumptions. Section~III formulates the long-term stochastic effective-EE maximization problem and proposed solution. Section~IV establishes convergence, conditional long-run performance,
physical battery safety, beam-pattern properties, robustness bounds, and computational complexity. Section~V presents the numerical results and discusses the resulting utility--battery-reserve trade-offs. Section~VI
concludes the paper.

\section{System and Channel Model}
\label{Section2}


We consider a downlink cell-free NTN comprising $M$ LEO satellites,
\rev{indexed by the satellite set $\mathcal{I}=\{1,\dots,M\}$, that jointly serve $U$ single-antenna UEs, indexed by the user set $\mathcal{J}=\{1,\dots,U\}$}, over a shared
time-frequency resource of bandwidth $W$. As illustrated in
Fig.~\ref{fig:system_model}, the satellites move through alternating sunlit
and eclipse portions of their orbits, exchange information through
inter-satellite links, and may access a GEO-assisted backhaul connection.
Each satellite is equipped with a uniform planar array (UPA) of
$N=N_x\times N_y$ elements and a hybrid analog--digital front end with
$N_{\mathrm{rf}}$ RF chains, where $N_{\mathrm{rf}}\ll N$. Under the
user-centric paradigm, UE $j$ is served by a dynamically formed satellite
cluster $\mathcal{S}_j\subseteq\mathcal{I}$; the set of UEs served by
satellite $i$ is
$\mathcal{U}_i=\{j\in\mathcal{J}\mid i\in\mathcal{S}_j\}$, with
$|\mathcal{U}_i|\le N_{\mathrm{rf}}$. The overlapping spot beams in
Fig.~\ref{fig:system_model} illustrate this cooperative transmission model,
including the example in which UE~3 is jointly served by LEOs~3 and~4.
Time is slotted into control intervals $t\in\{0,1,2,\dots\}$ of duration
$\Delta$, chosen on the order of the \emph{beam coherence time}, i.e., the
interval over which geometry-driven sCSI remains approximately valid. All
association and beamforming decisions are refreshed once per slot; within a
slot, the channel is treated as sCSI-stationary.


Satellite and UE positions in an Earth-centered, Earth-fixed frame are
denoted $\bm{q}_i(t) \in \R^3$ and $\bm{q}_j \in \R^3$, respectively, with
$\bm{q}_i(t)$ obtained from orbital ephemeris and $\bm{q}_j$ from GNSS-based
UE localization. The downlink channel vector from satellite $i$ to UE $j$
at slot $t$ follows a non-shadowed Rician model,
\begin{equation}
\bm{c}_{i,j}(t) = \sqrt{N\alpha_{i,j}}\left(
\sqrt{\tfrac{\varkappa_{i,j}}{\varkappa_{i,j}+1}}\, e^{j\varphi_{i,j}(t)}
+ \sqrt{\tfrac{1}{\varkappa_{i,j}+1}}\, x_{i,j}\right)\bm{e}_{i,j},
\label{eq:channel}
\end{equation}
where $\alpha_{i,j}$ is the large-scale fading coefficient (LSFC),
$\varkappa_{i,j}$ is the Rician $K$-factor, $x_{i,j}\sim\mathcal{CN}(0,1)$ is
the small-scale fading component, $\varphi_{i,j}(t)$ is the propagation phase at slot $t$ due to Doppler and propagation
delay $\tau_{i,j}$ at carrier frequency $f_c$, and $\bm{e}_{i,j} \triangleq
\bm{e}(\vartheta_{i,j},\phi_{i,j}) \in \C^N$ is the UPA steering vector at
zenith/azimuth angle of departure $(\vartheta_{i,j},\phi_{i,j})$, defined
analogously to a standard separable UPA response. The LSFC follows a
free-space-plus-atmospheric-attenuation model,
\begin{equation}
\alpha_{i,j} = \frac{G^{\mathrm{sat}}G^{\mathrm{ue}}\,\Lambda(\vartheta_{i,j})}
{(4\pi f_c \tau_{i,j})^2},
\label{eq:lsfc}
\end{equation}
with $G^{\mathrm{sat}}, G^{\mathrm{ue}}$ the satellite/UE antenna gains and
$\Lambda(\cdot)$ the zenith-angle-dependent atmospheric attenuation factor.
The tuple $\mathcal{H}_{i,j} \triangleq \{\alpha_{i,j}, \varkappa_{i,j},
f_{D,i,j}, \tau_{i,j}, \vartheta_{i,j}, \phi_{i,j}\}$ is the geometry-driven
sCSI parameter set, computable from $(\bm{q}_i(t),\bm{q}_j)$ alone and
therefore updatable at the (slow) control-slot rate $\Delta$ rather than at
the (fast) instantaneous fading coherence time.
The transmitted signal of satellite $i$ can be expressed as
\begin{equation}
\bm{x}_i(t) = \sum_{j \in \mathcal{U}_i} \bm{w}_{i,j}(t)\, d_j(t),
\end{equation}
where $d_j(t)\sim\mathcal{CN}(0,1)$ is the data symbol for UE $j$ and
$\bm{w}_{i,j}(t) = e^{j\varphi_{i,j}(t)}\bm{f}_{i,j}(t) \in \C^N$ is the
downlink beamformer, factored into a Doppler pre-compensation phase and a
composite hybrid beamformer $\bm{f}_{i,j}(t) = \bm{A}_i(t)\bm{b}_{i,j}(t)$,
with $\bm{A}_i(t)\in\C^{N\times N_{\mathrm{rf}}}$ a constant-modulus analog
precoder and $\bm{b}_{i,j}(t)\in\C^{N_{\mathrm{rf}}}$ the digital precoder.
Following channel hardening in the massive-antenna, user-centric cooperative
regime, each UE detects its signal using the statistical mean effective
channel; beamforming uncertainty and residual inter-user interference are
absorbed into an effective noise term. This yields the following SINR
characterization.

\begin{theorem}[SINR of a served UE]
\label{thm:sinr}
The downlink SINR of UE $j$ at slot $t$ is
\begin{equation}
\Gamma_j(t)=\frac{S_j(t)}{I_j(t)},
\label{eq:sinr}
\end{equation}
where
\revstart
\begin{align}
S_j(t)&\triangleq\left|\sum_{i\in\mathcal S_j}\sqrt{\varkappa_{i,j}}\,
\bm g_{i,j}^{\mathsf H}\bm f_{i,j}(t)\right|^2,\\
I_j(t)&\triangleq\sum_{n\in\mathcal J}\sum_{i\in\mathcal S_n}
\left|\bm g_{i,j}^{\mathsf H}\bm f_{i,n}(t)\right|^2\notag\\
&\quad+\sum_{n\in\mathcal J\setminus\{j\}}\sum_{p,r\in\mathcal S_n}
\bm f_{p,n}^{\mathsf H}(t)\bm D_{p,r,j,n}\bm f_{r,n}(t)+\sigma^2.
\end{align}
Let $\bm g_{i,j}=\sqrt{N\alpha_{i,j}/(\varkappa_{i,j}+1)}\,\bm e_{i,j}$ and
$\bm m_{i,j}=\sqrt{\varkappa_{i,j}}\,\bm g_{i,j}$. For a within-slot time
$u\in[0,\Delta]$, write $\delta_{i,j,n}(u)=\varphi_{i,j}(u)-\varphi_{i,n}(u)$,
where $\varphi_{i,k}(u)=2\pi(u f_{D,i,k}-f_c\tau_{i,k})$ in slot-local coordinates. Then
\begin{align}
\bm D_{p,r,j,n}
&\triangleq\frac1\Delta\int_0^\Delta
 e^{\mathrm i(\delta_{p,j,n}(u)-\delta_{r,j,n}(u))}
 \bm m_{p,j}\bm m_{r,j}^{\mathsf H}\,\mathrm du\notag\\
&=e^{\mathrm i\pi(\Delta\eta_{p,r,j,n}-2f_c\varepsilon_{p,r,j,n})}
\operatorname{sinc}(\Delta\eta_{p,r,j,n})
\bm m_{p,j}\bm m_{r,j}^{\mathsf H}.\label{7}
\end{align}
Here $\operatorname{sinc}(x)=\sin(\pi x)/(\pi x)$,
$\eta_{p,r,j,n}=f_{D,p,j}-f_{D,p,n}+f_{D,r,n}-f_{D,r,j}$, and
$\varepsilon_{p,r,j,n}=\tau_{p,j}-\tau_{p,n}+\tau_{r,n}-\tau_{r,j}$.
The use-and-then-forget spectral efficiency is $R_j(t)=\log_2(1+\Gamma_j(t))$.
\end{theorem}
\begin{proof}
Assume independent unit-variance data streams and zero-mean diffuse
coefficients uncorrelated across satellites for each receiving UE.
Precompensation cancels the desired LoS phases, yielding the coherent
coefficient in $S_j(t)$. The diffuse variances sum over all streams,
including stream $j$. For each $n\ne j$, the residual LoS amplitude
at UE $j$ is $\sum_{i\in\mathcal S_n}e^{-\mathrm i\delta_{i,j,n}(u)}
\bm m_{i,j}^{\mathsf H}\bm f_{i,n}(t)$. Averaging its squared magnitude
over the slot gives the block quadratic form in $I_j(t)$. The block
matrix is positive semidefinite because it is an average of outer
products. Adding the independent receiver noise completes the bound.
\end{proof}
\revstop


The total electrical load of satellite $i$ is decomposed into an exogenous
platform load and a communication load. Let $P_i^{\mathrm{bus}}(t)$ denote
attitude-control, thermal, telemetry, housekeeping, and other non-communication
loads that must be supplied by the same electrical power subsystem. The
communication load is
\begin{align}
P_i^{\mathrm{com}}(t)
&=P_i^{\mathrm{fix}}+e_i(t)+P_i^{\mathrm{fw}}W
\sum_{j\in\mathcal U_i(t)}R_j(t),\label{eq:pcom}\\
e_i(t)&\triangleq \frac{1}{\eta_i}\sum_{j\in\mathcal U_i(t)}
\|\bm f_{i,j}(t)\|_2^2,\label{eq:econtrol}
\end{align}
where $P_i^{\mathrm{fix}}=P_i^{\mathrm{bb}}+NP_i^{\mathrm{ant}}+
N_{\mathrm{rf}}P_i^{\mathrm{rf}}+P_i^{\mathrm{isl}}$, $\eta_i$ is the
power-amplifier efficiency, and $P_i^{\mathrm{fw}}$ is the forwarding-energy
coefficient. The battery-supplied load is therefore
\begin{equation}
P_i^{\mathrm{load}}(t)=P_i^{\mathrm{bus}}(t)+P_i^{\mathrm{com}}(t).
\label{eq:pload}
\end{equation}
This full-load model prevents the energy controller from declaring a
satellite sustainable while ignoring fixed platform and forwarding loads.
Next, we study the solar harvesting and physical battery dynamics. Let $\bm u_\odot(t)\in\mathbb R^3$ denote the ephemeris-computable Sun
direction. With eclipse indicator $\beta_i(t)\in\{0,1\}$, the harvested
power is
\begin{equation}
H_i(t)=\beta_i(t)\zeta_iA_iG_\odot[\cos\psi_i(t)]^+,
\label{eq:harvest}
\end{equation}
where $[x]^+=\max\{x,0\}$. The battery state $E_i(t)$ is measured at the
beginning of every slot and satisfies
\begin{equation}
E_i(t+1)=E_i(t)+\Delta\big(\theta_iH_i(t)-P_i^{\mathrm{load}}(t)\big),
\label{eq:battery}
\end{equation}
provided the controller enforces the physical bounds below. Harvested energy
is assumed usable during the current slot; when the power architecture only
permits next-slot use, $H_i(t)$ is omitted from the current-slot cap without
changing the subsequent analysis.

Energy causality and overflow avoidance are imposed \emph{robustly}, using
the controller's harvested-power estimate $\widehat H_i(t)$ together with
its assumed error bound $\delta_{H,i}$ (Assumption~\ref{as:causal}), rather
than the point estimate alone:
\begin{align}
\Delta P_i^{\mathrm{load}}(t)
&\le E_i(t)-E_i^{\min}\notag\\
&\quad+\Delta\theta_i\big[\widehat H_i(t)-\delta_{H,i}\big]^+,
\label{eq:causality}\\
E_i(t)-\Delta P_i^{\mathrm{load}}(t)
&\le E_i^{\max}\notag\\
&\quad-\Delta\theta_i\big[\widehat H_i(t)+\delta_{H,i}\big].
\label{eq:overflow}
\end{align}
The depletion constraint uses the lower harvest estimate and the overflow constraint uses the upper harvest estimate, so that both constraints hold for the true $H_i(t)$
regardless of which way the estimation error points. Using the point estimate without a robustness margin would guarantee safety only under exact harvest knowledge. The robust formulation instead ensures that clipping is not
used to mask an infeasible energy decision, and, as shown in
Corollary~\ref{cor:battery}, every implemented action now provably
preserves $E_i(t+1)\in[E_i^{\min},E_i^{\max}]$ under bounded harvest error,
not only under exact harvest knowledge.
In this work, we consider the following assumptions.
\begin{assumption}[Causal state information, with bounded harvest error]\label{as:causal}
At the beginning of slot $t$, the controller knows the measured battery
state $E_i(t)$, current geometry-driven sCSI, and a causal estimate
$\widehat H_i(t)$ of the \rev{average harvested power over the slot}, with
$|\widehat H_i(t)-H_i(t)|\le\delta_{H,i}$ for a known bound $\delta_{H,i}\ge0$.
No future realization is used in the control decision. Equation
\eqref{eq:battery}'s true battery evolution always uses the true,
retrospectively realized $H_i(t)$; only the controller's \emph{decision} at
slot $t$ is based on $\widehat H_i(t)$.
\end{assumption}

\begin{assumption}[Bounded exogenous processes]\label{as:ergodic}
The processes $H_i(t)$ and $P_i^{\mathrm{bus}}(t)$ are bounded and jointly
stationary and ergodic. Specifically, $0\le H_i(t)\le H_i^{\max}$ and
$0\le P_i^{\mathrm{bus}}(t)\le P_{i,\mathrm{bus}}^{\max}$.
\end{assumption}

\begin{assumption}[Per-slot feasibility]\label{as:qosfeas}
For every admissible state, the QoS, RF-chain, hybrid-beamforming, peak-power,
and physical battery constraints admit at least one feasible action. If this
condition is violated, admission control or temporary QoS relaxation must be
introduced; such mechanisms are outside the present scope.
\end{assumption}

\revstart
\begin{assumption}[Comparison policy]\label{as:slack}
On the states visited by the implemented controller, a causal reference
rule selects actions from the same instantaneous feasible set. Its utility
and queue-weighted net load are evaluated along this visited state process.
\end{assumption}
\revstop
\begin{assumption}[Bounded available-energy error]\label{as:predict}
Let $\rho_i(t)$ be the true communication-power budget induced by
\eqref{eq:causality}--\eqref{eq:overflow} evaluated at the true $H_i(t)$,
after subtracting the known platform and fixed communication loads, and let
$\widehat\rho_i(t)$ be the budget the controller would compute from the
point estimate $\widehat H_i(t)$ alone (i.e. without the robustness margin
$\delta_{H,i}$). Since $\rho_i(t)$ and $\widehat\rho_i(t)$ differ only
through $\theta_i(H_i(t)-\widehat H_i(t))$, Assumption~\ref{as:causal}
gives $|\widehat\rho_i(t)-\rho_i(t)|\le\theta_i\delta_{H,i}=:\delta_i$.
This assumption is stated directly on the usable power budget and
therefore does not incorrectly identify a one-slot harvested-power error
with a recursively accumulated battery-state error. It is used only in the
operating-point sensitivity analysis of Theorem~\ref{thm:sensitivity},
which quantifies the value variation of a fixed convex SCA subproblem when
its available-energy-budget vector is perturbed. It does not provide a
global sensitivity bound for the original nonconvex per-slot problem and
plays no role in the physical battery-safety guarantee of
Corollary~\ref{cor:battery}, which instead relies on the explicit margin
$\delta_{H,i}$ used in \eqref{eq:causality}--\eqref{eq:overflow}.
\end{assumption}


Total instantaneous network power is $P^{\mathrm{tot}}(t) =
\sum_{i\in\mathcal{I}}\big(P_i^{\mathrm{fix}} + e_i(t)\big) +
\sum_{i\in\mathcal{I}} P_i^{\mathrm{fw}} W \sum_{j\in\mathcal{U}_i}R_j(t)$,
and instantaneous network EE is
\begin{equation}
\mathrm{EE}(t) \triangleq \frac{W\sum_{j\in\mathcal{J}} R_j(t)}{P^{\mathrm{tot}}(t)}.
\end{equation}
As in static-power designs, the achievable rate appears in both numerator
and denominator (through the forwarding term), which complicates direct
optimization. The same monotonicity argument used for static-power EE
applies unchanged here (it does not involve the battery dynamics), giving a
tractable \emph{effective} EE surrogate.

\begin{lemma}[Effective EE lower bound]
\label{lem:eeslb}
$\mathrm{EE}(t)$ is lower bounded by a monotonically increasing function of
\begin{equation}
\Xi(t) \triangleq \frac{W\sum_{j\in\mathcal{J}} R_j(t)}
{\sum_{i\in\mathcal{I}}P_i^{\mathrm{fix}} + \sum_{i\in\mathcal{I}}e_i(t)}.
\end{equation}
Consequently, maximizing $\Xi(t)$ at every slot maximizes the lower bound on
$\mathrm{EE}(t)$.
\end{lemma}

\begin{proof}
Upper-bound the per-satellite forwarding term by
$P_i^{\mathrm{fw}}W\sum_{j}R_j(t) \le M\max_i P_i^{\mathrm{fw}}\,
W\sum_j R_j(t)$, substitute into $P^{\mathrm{tot}}(t)$, and divide numerator
and denominator of $\mathrm{EE}(t)$ by $\big(\sum_i P_i^{\mathrm{fix}} +
\sum_i e_i(t)\big)$ to obtain $\mathrm{EE}(t) \ge \Xi(t)\big/\big(1 +
M\max_iP_i^{\mathrm{fw}}\,\Xi(t)\big)$, a function that is increasing in
$\Xi(t)$ on $\Xi(t)\ge0$ (its derivative w.r.t. $\Xi(t)$ is
$1/(1+M\max_iP_i^{\mathrm{fw}}\Xi(t))^2>0$).
\end{proof}

We henceforth use $\Xi(t)$ as the per-slot objective, exactly as the
static-power baseline uses its analogous effective-EE surrogate.

\section{Problem Formulation and Proposed Solution}

\subsection{Problem Formulation}

\revstart
Let $\Phi(t)$ collect the clustering and hybrid-beamforming variables.
The maximization is over causal policies. For a given policy and initial
battery state, $\mathbb E$ averages over the exogenous random inputs
(including harvest and platform-load innovations where modeled) and
policy randomization. The battery distribution is induced by their history
and the battery recursion. Deterministic orbital inputs are conditioned
on, and the small-scale fading average is already included in the rate
bound. The limit inferior is applied to the sequence of finite-horizon
expected averages as $T$ grows.
The effective efficiency $\Xi(t)$ keeps the denominator quadratic in
the beamformers by removing rate-dependent forwarding power from the
objective. Forwarding remains in the actual electrical load. Lemma~1
relates $\Xi(t)$ to instantaneous EE. Over a horizon with equal-duration
slots, cumulative communication EE equals
$\sum_t P^{\mathrm{tot}}(t)\mathrm{EE}(t)/\sum_t P^{\mathrm{tot}}(t)$,
a power-weighted average of instantaneous EE. The two objectives may
rank policies differently.
\revstop
\begin{subequations}\label{eq:P0}
\begin{align}
\mathscr P_0:\quad \max_{\{\Phi(t)\}}\;&
\bar\Xi\triangleq\liminf_{T\to\infty}\frac1T
\sum_{t=0}^{T-1}\mathbb E[\Xi(t)]\label{eq:P0obj}\\
\mathrm{s.t.}\quad
&R_j(t)\ge R_j^{\min},\ \forall j,t,\label{eq:P0qos}\\
&|\mathcal U_i(t)|\le N_{\mathrm{rf}},\ \forall i,t,\label{eq:P0rf}\\
&e_i(t)\le e_i^{\max},\ \forall i,t,\label{eq:P0peak}\\
&\eqref{eq:causality},\ \eqref{eq:overflow},\ \forall i,t,\label{eq:P0power}\\
&\bm f_{i,j}(t)=\bm A_i(t)\bm b_{i,j}(t),\quad
\bm A_i(t)\in\mathcal A.\label{eq:P0hybrid}
\end{align}
\end{subequations}
Because \eqref{eq:battery} couples present actions to future feasible sets,
$\mathscr P_0$ is an infinite-horizon stochastic control problem.


Beyond the mixed-integer, nonconvex, fractional structure already present
in static-power cell-free EE design (discrete clustering $\mathcal{S}_j$,
nonconcave rate $R_j$ in $\bm{f}$, fractional objective $\Xi$, hybrid
beamforming constraint \eqref{eq:P0hybrid}), $\mathscr{P}_0$ additionally
exhibits \emph{temporal coupling} through \eqref{eq:battery}: the feasible
set of $\Phi(t)$ depends on $E_i(t)$, which depends on the entire past
trajectory $\{e_i(\tau)\}_{\tau<t}$. A myopic policy that simply resolves a
static-power problem each slot with $\rho_i(t)$ set to the instantaneously
available power ignores the opportunity cost of using battery energy now
versus preserving it for an upcoming eclipse, and provides no guarantee on
long-run average EE relative to a policy that plans across the battery
dynamics.

\subsection{Proposed Solution}\label{sec:solution}

\subsubsection{Perturbed Physical-Battery Queue and Lyapunov Drift}

Choose a perturbation level $\Omega_i\in(E_i^{\min},E_i^{\max})$ and define
\begin{equation}
Z_i(t)\triangleq \Omega_i-E_i(t).
\label{eq:vqueue}
\end{equation}
Using \eqref{eq:battery}, its exact evolution is
\begin{equation}
Z_i(t+1)=Z_i(t)+\Delta\big(P_i^{\mathrm{load}}(t)-\theta_iH_i(t)\big).
\label{eq:zupdate}
\end{equation}
Unlike an independent deficit queue, $Z_i(t)$ is an affine transformation
of the measured physical state of charge. Therefore, controlling $Z_i(t)$
and enforcing \eqref{eq:causality}--\eqref{eq:overflow} directly controls
the true battery.

Let $L(\bm Z(t))=\frac12\sum_iZ_i^2(t)$. Boundedness of the harvested and
load powers gives
\begin{equation}
\Delta_L(t)\le B+\Delta\sum_i Z_i(t)
\mathbb E[P_i^{\mathrm{load}}(t)-\theta_iH_i(t)\mid\mathcal F_t],
\label{eq:driftbound}
\end{equation}
\rev{Here $\mathcal F_t$ is the information available at the slot start and $\Delta_L(t)=\mathbb E[L(\bm Z(t+1))-L(\bm Z(t))\mid\mathcal F_t]$.} where $B=\frac{\Delta^2}{2}\sum_i(P_{i,\mathrm{load}}^{\max}+
\theta_iH_i^{\max})^2$ is finite.

\subsubsection{Drift-Plus-Penalty Per-Slot Problem}

Dropping terms independent of $\Phi(t)$, the controller solves
\begin{subequations}\label{eq:P1t}
\begin{align}
\mathscr P_1(t):\quad\max_{\Phi(t)}\;&
V\Xi(t)-\Delta\sum_i Z_i(t)P_i^{\mathrm{load}}(t)
\label{eq:P1tobj}\\
\mathrm{s.t.}\quad&\eqref{eq:P0qos},\eqref{eq:P0rf},
\eqref{eq:P0peak},\eqref{eq:P0power},\eqref{eq:P0hybrid}.
\label{eq:P1tpower}
\end{align}
\end{subequations}
\revstart
The measured battery summarizes the realized history of charging and
load, so the controller can price current energy use without estimating
a future state transition kernel or solving a finite-horizon forecast
problem. The queue weight shapes the objective, while the robust action
set enforces the battery interval at each step.
\revstop
When $E_i(t)<\Omega_i$, $Z_i(t)>0$ and the queue term penalizes power use;
when the battery is above the perturbation level, $Z_i(t)<0$ and the
controller is encouraged to exploit otherwise spillable solar energy. The
hard physical constraints remain active in both cases.

\subsubsection{Group-Sparse Association}

As in the static-power baseline, we relax the discrete clustering
$\mathcal{S}_j$ into a continuous group-sparse beamforming representation.
Define $\tilde{\bm{f}}_{i,j}(t) = \bm{f}_{i,j}(t)$ if $i\in\mathcal{S}_j(t)$
and $\bm{0}$ otherwise, so that
$\mathcal{S}_j(t)=\{i\mid\mathbb{1}_{\R^+}(\|\tilde{\bm{f}}_{i,j}(t)\|_2)=1\}$
and the RF-chain constraint \eqref{eq:P0rf} becomes $\sum_{j}
\mathbb{1}_{\R^+}(\|\tilde{\bm{f}}_{i,j}(t)\|_2)\le N_{\mathrm{rf}}$. We
approximate this with reweighted-$\ell_2$ relaxation: at outer iteration
$m$, weights $\omega_{i,j}^{(m)} = \big(\|\tilde{\bm{f}}_{i,j}^{(m-1)}(t)\|_2^2
+ \epsilon\big)^{-1}$ replace the indicator with
$\sum_j \omega_{i,j}^{(m)}\|\tilde{\bm{f}}_{i,j}(t)\|_2^2 \le N_{\mathrm{rf}}$,
exactly as in the static-power design; this component of the problem does
not interact with the battery dynamics and is therefore unchanged.

\subsubsection{Concave Surrogate Rate and Convex Reformulation}

The rate $R_j(t) = \log_2(1+\Gamma_j(t))$ is nonconcave in
$\tilde{\bm{f}}(t)$. We construct a concave quadratic surrogate via the
quadratic-transform technique.

\begin{proposition}[Concave surrogate lower bound]
\label{prop:surrogate}
For any $\gamma_j \ge 0$ and $\zeta_j\in\C$, define
\begin{align}
S_j(t) &\triangleq \Big|\sum_{i\in\mathcal S_j}\sqrt{\varkappa_{i,j}}\,
\bm{g}_{i,j}^{\mathsf H}\tilde{\bm f}_{i,j}(t)\Big|^2,\\
I_j(t)&\triangleq\sum_n\sum_{i\in\mathcal S_n}
\big|\bm g_{i,j}^{\mathsf H}\tilde{\bm f}_{i,n}(t)\big|^2\notag\\
&\quad+\sum_{n\in\mathcal J\setminus\{j\}}\sum_{p,r\in\mathcal S_n}
\tilde{\bm f}_{p,n}^{\mathsf H}(t)\bm D_{p,r,j,n}
\tilde{\bm f}_{r,n}(t)+\sigma^2,\\
h_j(\gamma_j,\zeta_j) &\triangleq -\log_2(1+\gamma_j) + \tfrac{1}{\ln2}
\rev{\gamma_j},\\
p_j\big(\tilde{\bm f},\gamma_j,\zeta_j\big)
&\triangleq\frac{2\sqrt{1+\gamma_j}}{\ln2}
\sum_{i\in\mathcal S_j}\sqrt{\varkappa_{i,j}}\notag\\
&\quad\times\operatorname{Re}\!\left\{\zeta_j^*\bm g_{i,j}^{\mathsf H}
\tilde{\bm f}_{i,j}\right\},\\
r_j\big(\tilde{\bm f},\zeta_j\big)
&\triangleq\rev{\frac{|\zeta_j|^2}{\ln2}\,(S_j+I_j)}.
\end{align}
Then the surrogate rate
$\hat R_j\big(\tilde{\bm f}(t),\gamma_j,\zeta_j\big) \triangleq
-h_j(\gamma_j,\zeta_j) + p_j\big(\tilde{\bm f}(t),\gamma_j,\zeta_j\big)
- r_j\big(\tilde{\bm f}(t),\zeta_j\big)$
satisfies $\hat R_j\big(\tilde{\bm f}(t),\gamma_j,\zeta_j\big) \le R_j(t)$
for all $\gamma_j,\zeta_j$, with equality when
$\gamma_j=\Gamma_j(t)$ and $\zeta_j = \sqrt{1+\gamma_j}\,S_j(t)^{1/2}\big/(S_j(t)+I_j(t))$
$\cdot\,\mathrm{sign}\!\big(\textstyle\sum_{i}\sqrt{\varkappa_{i,j}}\bm g_{i,j}^{\mathsf H}\tilde{\bm f}_{i,j}(t)\big)$,
the value that maximizes the right-hand side for fixed $\tilde{\bm f}(t)$.
Moreover, $\hat R_j$ is jointly concave in $\tilde{\bm f}(t)$ for fixed
$(\gamma_j,\zeta_j)$.
\end{proposition}

\begin{proof}
Concavity of $\log_2(1+x)$ gives the tangent-line bound $\log_2(1+x) \ge
\log_2(1+\gamma_j) - \tfrac{\gamma_j}{\ln2} +
\tfrac{(1+\gamma_j)x}{\ln2(1+x)}$ for $x=\Gamma_j(t)$; substituting
$\Gamma_j(t)=S_j(t)/I_j(t)$ and writing $x/(1+x)$ in terms of
$S_j(t),I_j(t)$ gives a bound in terms of the ratio $S_j(t)/(S_j(t)+I_j(t))$.
Applying the quadratic transform $-B|y-A/B|^2+A^2/B\le A^2/B$ with
$A=\sqrt{1+\gamma_j}\sum_i\sqrt{\varkappa_{i,j}}\bm g_{i,j}^{\mathsf H}
\tilde{\bm f}_{i,j}(t)$, $B = S_j(t)+I_j(t)$, and free variable
$y=\zeta_j$, and maximizing the resulting quadratic-over-linear surrogate
over $\zeta_j$ (attained at the stated closed form by setting the gradient
w.r.t. $\zeta_j$ to zero) recovers $-h_j+p_j-r_j$ as a lower bound on
$\log_2(1+\Gamma_j(t))$, tight at the stated $(\gamma_j,\zeta_j)$. Let $\bm D_{j,n}$ denote the block matrix whose $(p,r)$th block is
$\bm D_{p,r,j,n}$. Since $\bm D_{j,n}$ represents the covariance matrix
of the stacked cross-satellite channel coefficients, it is Hermitian
positive semidefinite. The closed-form blocks in \eqref{7} are therefore understood as
entries of this covariance matrix and hence preserve its Hermitian
positive-semidefinite structure.
Therefore, $r_j$ is a nonnegative-semidefinite quadratic form in
$\tilde{\bm f}(t)$, so $-r_j$ is concave. Since $p_j$ is affine,
$\hat R_j=-h_j+p_j-r_j$ is concave in $\tilde{\bm f}(t)$.
\end{proof}

Substituting $\hat R_j$ for $R_j$ in $\mathscr{P}_1(t)$, and alternating
between (i) updating $(\gamma_j,\zeta_j)$ in closed form at the current
iterate and (ii) re-optimizing $\tilde{\bm f}(t)$, produces a non-decreasing
sequence of per-slot objective values, exactly as in the static-power
design's inner loop.

\subsubsection{Quadratic-Transform and Sign-Aware SCA Solution}
\label{sec:qtsca}

Direct Charnes--Cooper scaling of the queue-penalized problem generates
terms proportional to $\|\bm v\|^2/\chi^2$, which are not jointly convex.
Moreover, applying a standard Dinkelbach iteration only to the EE ratio while
leaving the battery term additive is not exact for the objective in
\eqref{eq:P1tobj}. We therefore apply the scalar quadratic transform directly
to the EE term and treat the sign-dependent battery term by SCA.

For fixed rate auxiliaries, define
\begin{equation}
\widehat N(\tilde{\bm f},\bm t)\triangleq W\sum_j t_j,
\qquad
D(\tilde{\bm f})\triangleq P_0+\sum_i e_i(t),
\label{eq:NDdef}
\end{equation}
where $P_0=\sum_iP_i^{\mathrm{fix}}$, $t_j\le \widehat R_j$, and
$\widehat N\ge0$, $D>0$. The scalar quadratic-transform identity gives
\begin{equation}
V\frac{\widehat N}{D}
=
\max_{y\ge0}V\big(2y\sqrt{\widehat N}-y^2D\big),
\qquad
y^\star=\frac{\sqrt{\widehat N}}{D}.
\label{eq:eeqt}
\end{equation}
Introduce $s\ge0$ with $s^2\le W\sum_jt_j$, which is representable by a
rotated second-order cone. For fixed $y$, the transformed EE term
$2Vys-Vy^2D$ is concave.

\revstart
Let $\mathcal I_+=\{i:Z_i(t)\ge0\}$ and $\mathcal I_-=\{i:Z_i(t)<0\}$. The battery-weighted load and physical constraints require two-sided
bounds on forwarding load. Write $x$ for the real-stacked beamformers
and let $x^{(r)}$ be the current point. For the rate upper bound, let
$ell_j(x)$ be the affine tangent of the convex interference-plus-noise
quadratic $I_j(x)$ at $x^{(r)}$. On $\ell_j(x)\ge\sigma^2/2$, define
\begin{equation}
\overline R_j(x)=\log_2 C_j+
\frac{S_j(x)+I_j(x)-C_j}{C_j\ln 2}-\log_2\ell_j(x),
\end{equation}
where $C_j=S_j(x^{(r)})+I_j(x^{(r)})$. The tangent bound on the
first logarithm and $\ell_j\le I_j$ give
$\hat R_j\le R_j\le\overline R_j$. Both bounds match the true rate
and its gradient at the current iterate, and $\overline R_j$ is convex.
Let $P_i^+$ use exact PA power and $\overline R_j$ for forwarding.
Let $P_i^-$ use the affine PA-power tangent and $\hat R_j$ for
forwarding. Thus $P_i^-\le P_i^{\mathrm{load}}\le P_i^+$, with
$P_i^-$ concave and $P_i^+$ convex.
\begin{subequations}\label{eq:qtsca}
\begin{align}
\max_{x,\bm t,s}\quad
&2Vys-Vy^2D(x)-\Delta\sum_{i\in\mathcal I_+} Z_iP_i^+\notag\\
&\quad+\Delta\sum_{i\in\mathcal I_-}|Z_i|P_i^-
\label{eq:qtscaobj}\\
\mathrm{s.t.}\quad
&R_j^{\min}\le t_j\le\hat R_j(x),\quad
 \ell_j(x)\ge\sigma^2/2,\label{eq:qtscarate}\\
&0\le s,\quad s^2\le W\sum_j t_j,\label{eq:qtscasoc}\\
&P_i^+\le P_{i,\mathrm{cap}},\quad P_i^-\ge P_{i,\mathrm{floor}},
\label{eq:qtscaconst}
\end{align}
\end{subequations}
with the PA-power and fixed-weight sparse constraints also enforced.
Here $P_{i,\mathrm{cap}}$ and $P_{i,\mathrm{floor}}$ are the upper and
lower load limits in \eqref{eq:causality} and \eqref{eq:overflow}.
This construction enforces QoS, energy causality and overflow limits
for the current continuous relaxation. The lower load surrogate is
the sign-aware minorizer, denoted by $P_i^-$ in \eqref{eq:qtscaobj}.
\revstop
The objective in \eqref{eq:qtscaobj} is concave and the feasible set is
convex, provided the stacked interference covariance matrices in the rate
surrogate are Hermitian positive semidefinite. Hence \eqref{eq:qtsca} is a
convex program with a second-order cone and, when forwarding is modeled, logarithmic terms that admit exponential-cone representations. After each beamforming update, set
\begin{equation}
y^{(r+1)}=
\frac{\sqrt{W\sum_jt_j^{(r+1)}}}
{P_0+\sum_ie_i^{(r+1)}(t)},
\label{eq:yupdate}
\end{equation}
update $(\gamma_j,\zeta_j)$ to their tight values, rebuild the load
minorizer, and repeat. This procedure optimizes a globally lower-bounding,
first-order-consistent surrogate of the actual queue-penalized per-slot
objective and does not invoke standard Dinkelbach theory for a ratio-plus-additive-term objective.

\subsubsection{Hybrid Beamforming Recovery}

Given the fully-digital solution $\{\bm f_{i,j}^\star(t)\}$ obtained from
the convex quadratic-transform/SCA subproblem \eqref{eq:qtsca}, the constant-modulus analog
precoder $\bm A_i(t)$ and digital precoder $\bm B_i(t)$ are recovered per
satellite by projecting onto the hybrid feasible set,
\begin{equation}
\min_{\bm A_i\in\mathcal A,\ \bm B_i}\ \big\|\bm F_i^\star(t) - \bm A_i\bm B_i\big\|_F^2,
\end{equation}
solved by alternating a closed-form least-squares update of $\bm B_i$ (for
fixed $\bm A_i$) with a Riemannian conjugate-gradient step for $\bm A_i$ on
the complex-circle manifold $\mathcal A$ (constant-modulus constraint) --
an unmodified, standard technique from static-power hybrid-beamforming
designs, restated here only for completeness and omitted in detail as it
does not interact with the energy-harvesting extension. Algorithm~\ref{alg:main}
summarizes the complete per-slot procedure, including this recovery step
and the subsequent battery update.

\begin{algorithm}[t]
\caption{Battery-Aware User-Centric Clustering and Beamforming}
\label{alg:main}
\begin{algorithmic}[1]
\State \textbf{Input:} $V$, $\{\Omega_i\}$, measured $\{E_i(0)\}$, tolerances.
\For{$t=0,1,2,\ldots$}
\State Measure $E_i(t)$ and update sCSI and $\widehat H_i(t)$.
\State Set $Z_i(t)=\Omega_i-E_i(t)$ and form the physical power bounds.
\State Initialize feasible beamformers, sparse weights, and $y^{(0)}$.
\Repeat
\State Update $(\gamma_j,\zeta_j)$ at the current beamformers.
\State \rev{Form the queue sign sets and build the upper and lower load bounds.}
\State Solve the convex quadratic-transform/SCA subproblem \eqref{eq:qtsca}.
\State Update $y$ using \eqref{eq:yupdate} and update sparse weights.
\Until{all block residuals satisfy their stopping tolerances}
\State \rev{Recover clusters and hybrid beamformers. Apply only an action verified against the original constraints, using a separately available feasible fallback if needed.}
\State \rev{Use realized $H_i(t)$ for the physical balance \eqref{eq:battery}. Read measured $E_i(t+1)$ for the next decision.}
\EndFor
\end{algorithmic}
\end{algorithm}

\section{Performance Analysis}

This section establishes convergence of the per-slot procedure, conditional long-run Lyapunov performance, physical battery safety, a conditional steering-projection result, separate sensitivity results for energy-budget and positioning errors, and computational complexity.

\subsection{Convergence of the Per-Slot Iterative Procedure}

\begin{proposition}[Monotonic convergence of the relaxed per-slot procedure]
\label{prop:conv}
\revstart
Assume that every conic subproblem \eqref{eq:qtsca} is solved exactly, the
relaxed feasible set is compact, and all rate and load surrogates are
value-tight and first-order tight at the current iterate, with the objective minorized and the feasible set approximated from within. For fixed sparse weights, the sequence of true queue-penalized
per-slot objective values generated by the rate-auxiliary, scalar
quadratic-transform, and sign-aware SCA updates is nondecreasing and
convergent. If the surrogate feasible-set map is inner semicontinuous
at each accumulation point and a constraint qualification holds for
the fixed-point subproblem there, each accumulation point is stationary
for the fixed-weight continuous relaxation.
\revstop
\end{proposition}

\begin{proof}
Let $\Psi(x)$ denote the actual relaxed objective in \eqref{eq:P1tobj} and
$\widehat\Psi(x\mid x^{(r)})$ the surrogate obtained from the tight rate
minorizer, the exact scalar quadratic transform at the updated auxiliary
$y^{(r)}$, and the affine load minorizer \eqref{eq:qtscaobj}. These
constructions satisfy
\begin{equation}
\widehat\Psi(x\mid x^{(r)})\le\Psi(x),\qquad
\widehat\Psi(x^{(r)}\mid x^{(r)})=\Psi(x^{(r)}),
\label{eq:mmprops}
\end{equation}
with matching first derivatives at $x^{(r)}$. Exact solution of
\eqref{eq:qtsca} therefore yields
\begin{align}
\Psi(x^{(r+1)})
&\ge \widehat\Psi(x^{(r+1)}\mid x^{(r)})\notag\\
&\ge \widehat\Psi(x^{(r)}\mid x^{(r)})
=\Psi(x^{(r)}).
\label{eq:mmchain}
\end{align}
\revstart
Compactness bounds the objective sequence. Let $x^\ast$ be an
accumulation point. If an admissible point for the surrogate at
$x^\ast$ strictly improved its surrogate value, inner semicontinuity
would supply nearby admissible points along the convergent subsequence.
By continuity and exact subproblem maximization, the objective gain
would then be bounded away from zero on that subsequence, contradicting
convergence. Hence $x^\ast$ maximizes its own surrogate. Its KKT
conditions transfer to the relaxation by first-order consistency.
Reweighting changes the feasible set, so the conclusion applies within
a fixed-weight stage. Cluster pruning and hybrid recovery require a
separate feasibility check.
\revstop
\end{proof}

\subsection{Long-Run Optimality and Physical Battery Safety}

\revstart
\begin{theorem}[Conditional Lyapunov performance bound]\label{thm:tradeoff}
Let $\Psi_t$ denote the objective in \eqref{eq:P1tobj}. Assume the
implemented hybrid action is feasible and its gap to the global supremum
of $\Psi_t$ on the same feasible set is $\epsilon_t\ge0$.
For a reference rule satisfying Assumption~\ref{as:slack}, define
\begin{align*}
\bar\Xi_{\mathrm{ref}}&=\liminf_{T\to\infty}
 T^{-1}\sum_{t<T}\mathbb E[\Xi_{\mathrm{ref}}(t)],\\
\bar D_{\mathrm{ref}}&=\limsup_{T\to\infty}T^{-1}\sum_{t<T}
 \Delta\sum_i\mathbb E[Z_i(t)(P_{i,\mathrm{ref}}^{\mathrm{load}}(t)-\theta_iH_i(t))],\\
\bar\epsilon&=\limsup_{T\to\infty}T^{-1}\sum_{t<T}\mathbb E[\epsilon_t].
\end{align*}
Then the implemented policy satisfies
\begin{equation}
\bar\Xi\ge\bar\Xi_{\mathrm{ref}}-
 \frac{B+\bar\epsilon+\bar D_{\mathrm{ref}}}{V}.
\label{eq:tradeoff}
\end{equation}
If $\bar D_{\mathrm{ref}}\le0$ and $\epsilon_t\le\epsilon_{\mathrm{alg}}$,
the loss relative to this reference is at most $(B+\epsilon_{\mathrm{alg}})/V$.
The $1/V$ scaling requires the numerator to remain bounded over the
considered family of designs.
\end{theorem}
\begin{proof}
Subtract $V\mathbb E[\Xi(t)\mid\mathcal F_t]$ from
\eqref{eq:driftbound}. The action-dependent terms are
$-\mathbb E[\Psi_t\mid\mathcal F_t]$. By the per-slot gap definition,
these are at most the corresponding reference terms plus
$\mathbb E[\epsilon_t\mid\mathcal F_t]$. The harvest contribution is
common to the two actions. Take expectations, sum over $t<T$ and
telescope the drift. Since $L(\bm Z(T))\ge0$, the only initial-state
term in the resulting lower bound is $-\mathbb E[L(\bm Z(0))]/(VT)$.
Taking the lower limit gives \eqref{eq:tradeoff}.
\end{proof}
The gap is assessed after cluster selection and hybrid recovery. On
small cases, a certified upper bound $U_t$ on the original feasible
optimum gives $\epsilon_t\le U_t-\Psi_t(\Phi(t))$. A local relaxed
iterate does not supply this certificate. Empirical checks over sampled
states characterize those states alone. A reference with conditional
mean net load zero on the visited states has $\bar D_{\mathrm{ref}}=0$.
\revstop
\begin{corollary}[Physical battery safety under bounded harvest error]\label{cor:battery}
If $E_i(0)\in[E_i^{\min},E_i^{\max}]$, Assumption~\ref{as:causal} holds, and
every implemented action satisfies the robust constraints
\eqref{eq:causality}--\eqref{eq:overflow}, then $E_i(t)\in[E_i^{\min},E_i^{\max}]$
for every $t$, regardless of the sign or realization of the harvest-prediction
error, provided only that it stays within the assumed bound $\delta_{H,i}$.
\end{corollary}

\begin{proof}
By induction on $t$. The base case holds by hypothesis. Suppose
$E_i(t)\in[E_i^{\min},E_i^{\max}]$. Since $H_i(t)\ge\widehat H_i(t)-\delta_{H,i}$
(Assumption~\ref{as:causal}) and $H_i(t)\ge0$, we have
$\theta_iH_i(t)\ge\theta_i[\widehat H_i(t)-\delta_{H,i}]^+$; combined with
\eqref{eq:causality} and the true update \eqref{eq:battery}, this gives
$E_i(t+1)=E_i(t)+\Delta(\theta_iH_i(t)-P_i^{\mathrm{load}}(t))\ge
E_i(t)+\Delta\theta_i[\widehat H_i(t)-\delta_{H,i}]^+-\Delta P_i^{\mathrm{load}}(t)
\ge E_i^{\min}$. Similarly $H_i(t)\le\widehat H_i(t)+\delta_{H,i}$ together
with \eqref{eq:overflow} gives $E_i(t+1)\le E_i(t)+\Delta(\theta_i[\widehat
H_i(t)+\delta_{H,i}]-P_i^{\mathrm{load}}(t))\le E_i^{\max}$. Hence
$E_i(t+1)\in[E_i^{\min},E_i^{\max}]$, completing the induction. The
Lyapunov result of Theorem~\ref{thm:tradeoff} controls the
reserve/optimality trade-off; per-slot battery safety, including under
bounded harvest-prediction error, follows from the physical, robustly
formulated action constraints rather than from mean-rate stability of an
unrelated virtual queue.
\end{proof}

\revstart
\begin{remark}
For fixed hardware, Corollary~\ref{cor:battery} directly yields
$|Z_i(t)|\le\max\{\Omega_i-E_i^{\min},E_i^{\max}-\Omega_i\}$ uniformly
in time and in $V$ for feasible actions. A battery sizing law as a
function of $V$ would require a separate family of designs and a
feasibility argument for that family.
\end{remark}
\revstop
\subsection{Beam Alignment in an Interference-Free Special Case}
\revstart
\begin{proposition}[Feasible steering projection in a battery-deficit slot]
\label{prop:align}
Fix the clusters and suppose inter-user interference is absent in the
rate model. Let $Z_i(t)\ge0$ for all satellites and let the steering
vectors have unit norm. For any feasible hybrid action, consider the
simultaneous projection
\begin{equation}
\widetilde{\bm f}_{i,j}=
\bm e_{i,j}\big(\bm e_{i,j}^{\mathsf H}\bm f_{i,j}\big).
\label{eq:aligned}
\end{equation}
If the projected beams admit a feasible hybrid factorization and each
projected load remains above the robust minimum load, the projection
weakly improves the slot objective. In particular, an aligned optimum
exists if an optimal action satisfies these projection conditions.
\end{proposition}
\begin{proof}
The projection preserves the complex desired-link inner products,
including their phases. In the interference-free model these inner
products determine the desired signal and its diffuse variance, so
all rates and forwarding loads are preserved. Orthogonal projection
weakly decreases each beam norm. The PA-power caps and the robust
maximum-load constraints thus remain satisfied. The hypothesis
preserves the minimum-load and hybrid constraints, and the fixed
clusters preserve the RF-chain count. The numerator of $\Xi(t)$ is
unchanged while its denominator weakly decreases. The queue-weighted
load term also weakly improves because each $Z_i(t)$ is nonnegative.
\end{proof}
The minimum-load condition matters near the battery ceiling because
lowering power can make overflow avoidance infeasible. This result
applies only under the stated projection conditions. It does not
establish an asymptotic optimality gap for the interference-coupled
problem. The snapshot in Section~\ref{sec:matrix} illustrates a
separate finite-array experiment with interference.
\revstop

\subsection{Sensitivity to Harvest-Prediction and
Available-Energy-Budget Error}
\label{subsec:budget_sensitivity}

For a fixed SCA iteration $r$, let
$\widehat{\Psi}^{(r)\star}(t;\bm{\rho})$ denote the optimal value of the
convex SCA subproblem in \eqref{eq:qtsca} under the
communication-power budget vector $\bm{\rho}$. The rate auxiliaries,
quadratic-transform variable, sparse weights, and load-linearization point
are held fixed at their values in iteration $r$. Let
$\widehat{\bm{\lambda}}^{(r)\star}\succeq\bm{0}$ denote the optimal dual
multiplier vector associated with the corresponding communication-power
constraints when the subproblem is solved at the estimated operating point
$\widehat{\bm{\rho}}$.

\begin{theorem}[Operating-point sensitivity of the convex SCA subproblem]
\label{thm:sensitivity}
Suppose that the fixed convex SCA subproblem satisfies Slater's condition
and that
\begin{equation}
\left|\widehat{\rho}_i-\rho_i\right|
\leq \delta_i,
\qquad \forall i.
\label{eq:budget_error_componentwise}
\end{equation}
Then,
\begin{equation}
\widehat{\Psi}^{(r)\star}(t;\bm{\rho})
-
\widehat{\Psi}^{(r)\star}
(t;\widehat{\bm{\rho}})
\leq
\sum_i
\widehat{\lambda}^{(r)\star}_i\delta_i.
\label{eq:sensitivity}
\end{equation}
\end{theorem}

\begin{proof}
For fixed SCA auxiliaries, sparse weights, and linearization point, the
subproblem is convex. Its optimal-value function is concave and
nondecreasing with respect to the right-hand-side resource vector
$\bm{\rho}$. Under Slater's condition, strong duality holds, and the
optimal multiplier vector
$\widehat{\bm{\lambda}}^{(r)\star}$ evaluated at
$\widehat{\bm{\rho}}$ is a supergradient of the optimal-value function.
Therefore,
\begin{align}
\widehat{\Psi}^{(r)\star}(t;\bm{\rho})
\leq\;&
\widehat{\Psi}^{(r)\star}
(t;\widehat{\bm{\rho}})
\nonumber\\
&+
\left(
\widehat{\bm{\lambda}}^{(r)\star}
\right)^{\mathsf T}
\left(
\bm{\rho}-\widehat{\bm{\rho}}
\right).
\label{eq:sensitivity_supergradient}
\end{align}
Since
$\widehat{\bm{\lambda}}^{(r)\star}\succeq\bm{0}$ and
\[
\rho_i-\widehat{\rho}_i
\leq
\left|\rho_i-\widehat{\rho}_i\right|
\leq
\delta_i,
\]
we obtain
\begin{equation}
\left(
\widehat{\bm{\lambda}}^{(r)\star}
\right)^{\mathsf T}
\left(
\bm{\rho}-\widehat{\bm{\rho}}
\right)
\leq
\sum_i
\widehat{\lambda}^{(r)\star}_i\delta_i,
\end{equation}
which proves \eqref{eq:sensitivity}.
\end{proof}

\begin{remark}
Equation \eqref{eq:sensitivity} characterizes the operating-point value
variation of the fixed convex SCA subproblem. It is not a global
sensitivity bound for the original nonconvex queue-penalized per-slot
problem, because the rate auxiliaries, quadratic-transform variable,
sparse weights, and load-linearization point are held fixed in the
analysis. Moreover, the bound applies to the complete convexified
Lyapunov objective rather than to the effective EE alone. Deriving an
EE-only sensitivity bound would require an additional decomposition.
The multiplier
$\widehat{\bm{\lambda}}^{(r)\star}$ is obtained directly from the convex
subproblem solved at $\widehat{\bm{\rho}}$ and does not require an
unobservable multiplier associated with the true budget vector.
\end{remark}

\subsection{Sensitivity to Satellite Positioning Error}

\revstart
\begin{theorem}[Fixed-action position sensitivity]\label{thm:possens}
Fix the clusters and beamformers and suppose the channel quantities are
continuously differentiable on a neighborhood of the operating geometry.
For positive $\Xi(t)$ and small $\max_i\|\Delta\bm q_i\|_2\le\sigma_p$,
\begin{equation}
\frac{|\Delta\Xi(t)|}{\Xi(t)}
\le \sigma_p\,
\frac{\sum_{i,j}\|\nabla_{\bm q_i}R_j(t)\|_2}{\sum_j R_j(t)}
+o(\sigma_p).
\label{eq:possens}
\end{equation}
The gradients include both the desired power and the full interference
denominator.
\end{theorem}
\begin{proof}
By $R_j=[\ln(S_j+I_j)-\ln I_j]/\ln2$, the gradient is
$[(\nabla S_j+\nabla I_j)/(S_j+I_j)-\nabla I_j/I_j]/\ln2$.
The denominator of $\Xi$ is fixed when the action is held fixed.
Taylor expansion and the triangle inequality give \eqref{eq:possens}.
\end{proof}
This certificate concerns the achieved efficiency at a fixed action.
It does not bound the change in the optimizer, clustering decisions,
or the long-run value.
\revstop
\begin{table*}[t]
\centering
\caption{Simulation parameters}
\label{tab:params}
\begin{tabular}{@{}llll@{}}
\toprule
Parameter & Symbol & Value & Notes \\
\midrule
Satellites & $M$ & 8 & \\
Orbital period & $T_{\rm orb}$ & 5700~s ($\sim$95 min) & 600-km-class LEO \\
Slot duration & $\Delta$ & 30~s & 190 slots/orbit \\
Eclipse fraction & -- & 35\% & of orbital period \\
Peak harvest power & $H_i^{\max}$ & $U[370,430]$~W & per satellite, randomized \\
Charge efficiency & $\theta_i$ & 0.90 & \\
Platform bus load & $P_i^{\mathrm{bus}}$ & $U[36,44]$~W (mean) & $\pm3$~W sinusoidal, $\pm1$~W noise \\
\rev{Fixed comm.\ load} & $\rev{P_i^{\mathrm{fix}}}$ & $U[22,28]$~W & \\
UEs served per satellite & $n_i$ & $U\{2,3,4\}$ & fixed for the run \\
Rate-power proxy coeff. & $\gamma_i$ & $U[0.9,1.5]$~W$^{-1}$ & Eq.~\eqref{eq:rateproxy} \\
Effective bandwidth const. & $W_c$ & 20 & Mbit/s scale \\
RF hardware peak power & $P^{\mathrm{comm,peak}}$ & 60~W & per satellite \\
Battery capacity & $E^{\max}$ & \rev{300~Wh} & \\
\rev{State-of-charge floor} & $E^{\min}$ & 60~Wh (20\%) & \\
Lyapunov target & $\Omega_i$ & $0.6E^{\max}+0.4E^{\min}$ & 204~Wh \\
Queue scaling & $Q_{\rm scale}$ & $1\times10^5$~J & design knob, calibrated \\
Initial battery (nominal) & $E_i(0)$ & $0.70E^{\max}$ & Results 1,2,3,5,7,8 \\
Initial battery (stressed) & $E_i(0)$ & $0.50E^{\max}$ & Results 4,6 \\
Stressed-scenario harvest scale & -- & $0.70\times$nominal & Results 4,6 \\
Near-depletion threshold & -- & $(1+0.1)E^{\min}=66$~Wh & Result 6 \\
Fixed-power-budget sweep & $\rho$ & $\{20,35,50\}$~W & Result 6 \\
Equal-power constant & $\rho_{\mathrm{eq}}$ & $53.6$~W & calibrated, Result 6 \\
Monte Carlo trials & -- & 10 (200 for Thm.~\ref{thm:sensitivity} check) & see Sec.~\ref{sec:numerical} \\
Solver outer iterations & -- & $\le6$ & quadratic-transform $y$-update \\
Solver convergence tol. & -- & $10^{-6}$ relative & true objective value \\
\bottomrule
\end{tabular}
\end{table*}

\subsection{Scope of the Robustness Results}
\label{sec:unified}
The budget-error result in Theorem~\ref{thm:sensitivity} applies to the
optimal value of a fixed convex SCA subproblem, whereas
Theorem~\ref{thm:possens} gives a first-order perturbation bound for the
achieved effective EE under positioning error. Because the two results
concern different quantities and approximation levels, they are reported
separately rather than combined into a single long-run robustness
guarantee. A unified theorem would require a common value function and
uniform error bounds over all battery states.

\subsection{Complexity}

\revstart
Let $L_c$ be the number of candidate satellite--UE links and $d=2NL_c$
the number of real beamforming coordinates. In particular, the
candidate count may be as large as $MU$ before sparsification. Each
subproblem also has $U$ rate variables and a scalar quadratic-transform
epigraph. Factoring a dense $d\times d$ rate quadratic once costs
$O(d^3)$, and evaluating $U$ such dense quadratic forms costs
$O(Ud^2)$ per inner update. Structured block factors can reduce these
counts. A lifted conic implementation has a cone dimension that depends
on the ranks of the rate and load quadratic factors, plus $O(U)$
exponential cones for the logarithms if the forwarding upper bounds
are active in the model. If $K$ is the dimension of the assembled
primal--dual Newton system including these lifts and $J$ is the
interior-point iteration count, dense factorization contributes
$O(JK^3)$ time and $O(K^2)$ memory per solve. The slot uses at most
$I_{\mathrm{out}}I_{\mathrm{in}}$ such solves under the stated outer
and inner iteration caps. Reweighting takes $O(NL_c)$ per outer update
and the battery accounting takes $O(M)$ per slot.
For satellite $i$ with $k_i$ retained streams and $r=N_{\mathrm{rf}}$,
a dense least-squares hybrid update costs
$O(Nr^2+r^3+Nrk_i+r^2k_i)$, while a gradient evaluation for its analog
factor costs $O(Nrk_i)$. These costs are multiplied by the recovery
iteration and line-search counts and summed over satellites.
\revstop
\section{Numerical Results and Discussion}
\label{sec:numerical}

This section evaluates the control-layer behavior of the proposed framework. The implementation uses the scalar quadratic transform of Section~\ref{sec:qtsca}, propagates position errors through physical geometry, checks the dual-multiplier sensitivity certificate, and reports utility and battery-reserve metrics separately. Trajectory results are averaged over 10 Monte Carlo trials per operating point, while the lower-cost sensitivity experiment uses 200 independently sampled operating points. The reported confidence intervals should therefore be interpreted in light of the more limited trajectory-level sample size.

\begin{figure*}[t]
\centering

\includegraphics[width=0.96\linewidth]{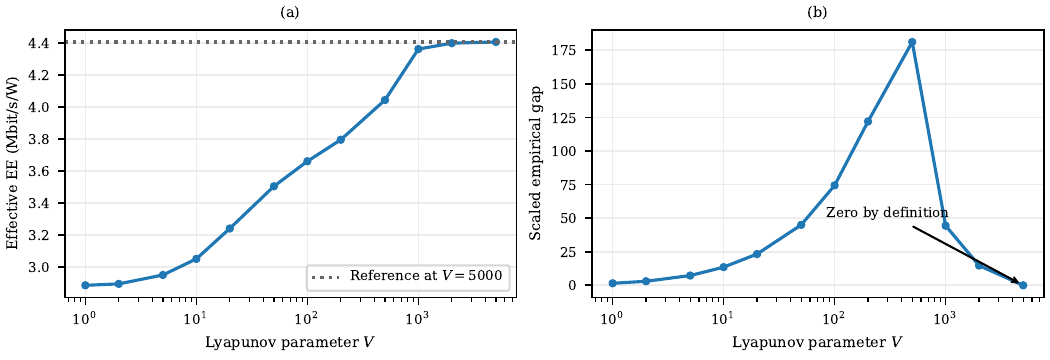}
\caption{\rev{(a) Mean effective EE over ten trials versus $V$.
(b) Scaled gap $V[\bar\Xi(5000)-\bar\Xi(V)]$ to the empirical endpoint.
The gap at $V=5000$ is zero by construction.}}
\label{fig:ee_vs_V}
\end{figure*}

\subsection{Control-Layer Evaluation Model}
\label{sec:scope}
The numerical study focuses on the control-layer behavior of the proposed energy-harvesting-aware framework. To enable long-horizon Monte Carlo evaluation over multiple orbital cycles and parameter sweeps, the matrix-level hybrid-beamforming and group-sparse clustering problem is represented by the calibrated per-satellite concave rate--power model
\begin{equation}
r_i(e_i)=W_c n_i\log_2\!\left(1+\frac{\gamma_i e_i}{n_i}\right).\label{eq:rateproxy}
\end{equation}
This surrogate preserves the monotonicity and concavity required by the per-slot energy-allocation controller and implements the same scalar quadratic-transform structure used in the proposed solution.

\revstart
The long-horizon experiments evaluate the reduced-order controller,
including its battery trajectories, uncertainty response, and policy
comparisons. The matrix snapshot below tests a separate fixed-cluster
special case.
\revstop
\revstart
NASA's small-spacecraft power survey documents multijunction photovoltaic
arrays and lithium-ion storage~\cite{nasa2026power}. At a nominal incident
flux of $1361$~W/m$^2$, a $1$~m$^2$ illuminated area at $30\%$ conversion
gives about $408$~W before incidence and system losses. A circular orbit
near $600$~km gives a Keplerian period near $97$~minutes. The $95$-minute
period, $35\%$ eclipse fraction, battery capacity and reserve floor, bus
and RF loads, and slot duration in Table~\ref{tab:params} define a
representative simulation scenario. They are scenario inputs rather than
specifications of a particular spacecraft.
\revstop
\subsection{Simulation Setup}
\label{sec:setup}
\revstart
The scalar simulator omits forwarding and uses an upper cap on
controllable power computed from the lower harvest estimate. It
clips the state after applying the realized harvest and load.
Upper clipping represents discarded charge. Lower clipping records
the battery at its prescribed floor even if the unprojected balance
would fall below it, which indicates a fixed-load energy shortfall.
The reported battery minima and near-depletion rates therefore
describe the clipped scalar model. A reproducibility check over
three stressed orbits at $V=20$ (one trial) found $0$, $86$, and
$1129$ below-floor satellite-slot updates before clipping for
the proposed, myopic, and fixed-$50$~W policies, respectively.
These counts diagnose the energy shortfall in that run.
Corollary~\ref{cor:battery}
applies separately to actions satisfying both robust constraints
before the battery update.
\revstop

Table~\ref{tab:params} lists every simulation parameter. In brief: $M=8$
LEO satellites on a common $\sim$95-minute ($T_{\rm orb}=5700$~s),
600-km-class orbit with control slots of $\Delta=30$~s ($190$
slots/orbit); $35\%$ eclipse fraction; solar panels sized to $\sim$1~m$^2$
at $30\%$ efficiency (peak harvest $370$--$430$~W per satellite,
randomized); battery capacity $E^{\max}=300$~Wh with a $20\%$
state-of-charge floor ($E^{\min}=60$~Wh); RF hardware peak-power
ceiling $60$~W per satellite. Under this calibration the time-averaged
harvested power exceeds the fixed platform/communication load by roughly
$27$--$44$~W per satellite in the nominal scenario, leaving a bounded
margin for communication power.
\begin{figure}[t]
\centering
\includegraphics[width=0.85\linewidth]{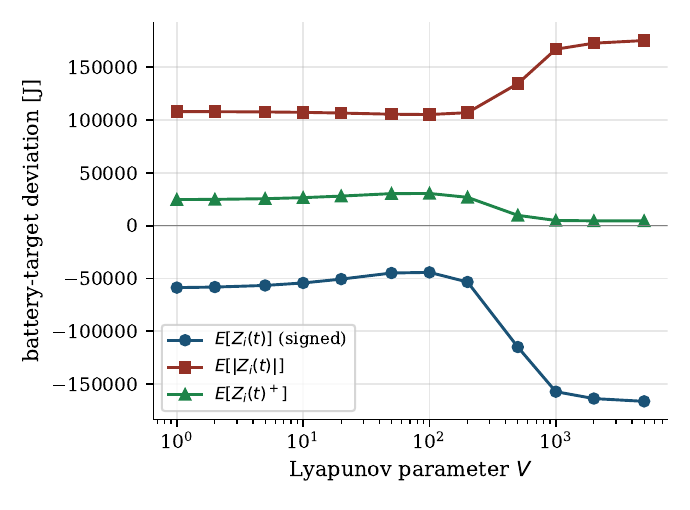}
\caption{Battery-target deviation metrics versus $V$: signed
$E[Z_i(t)]$, $E[|Z_i(t)|]$, and deficit-only $E[Z_i(t)^+]$, averaged over
10 trials.}
\label{fig:battery_vs_V}
\end{figure}

The per-slot solver alternates a closed-form $y$-update
(Eq.~\eqref{eq:yupdate}, applied to the surrogate) with a numerical solve
of the resulting concave subproblem over $\bm e$ (projected-gradient
ascent with a Barzilai--Borwein step size and a monotone backtracking
safeguard, since the subproblem is concave and a correct ascent step must
never decrease the objective), for up to 6 outer iterations or until the
true objective value changes by less than $10^{-6}$ relatively. All
robust-causality margins use $\delta_{H,i}=(\delta_H/H_i^{\max})H_i^{\max}$
with the tested $\delta_H/H^{\max}$ bound stated per experiment.

\begin{figure}[t]
\centering
\includegraphics[width=\linewidth]{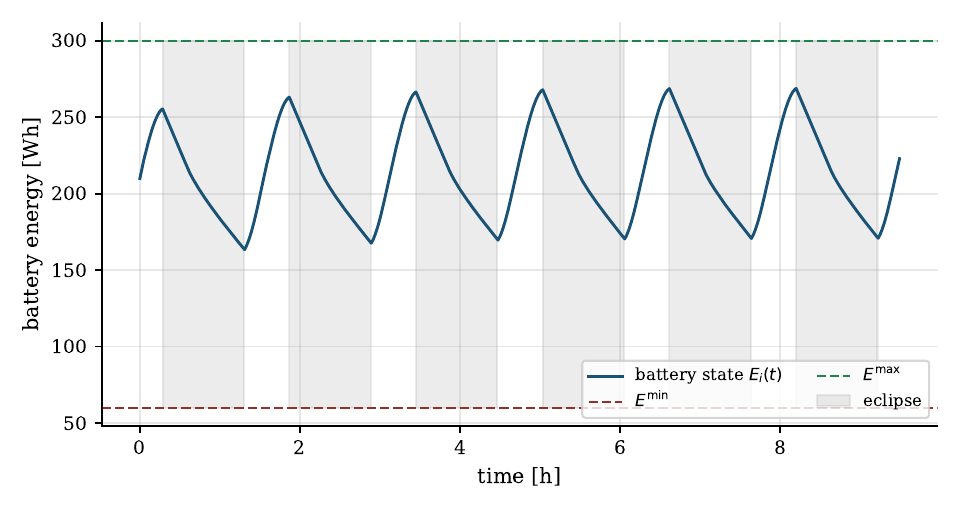}
\caption{Battery state-of-charge trajectory for a representative satellite
over 6 orbits at $V=100$; shaded bands mark eclipse periods.}
\label{fig:trajectory}
\end{figure}
\begin{figure*}[t]
\centering
\includegraphics[width=0.95\linewidth]{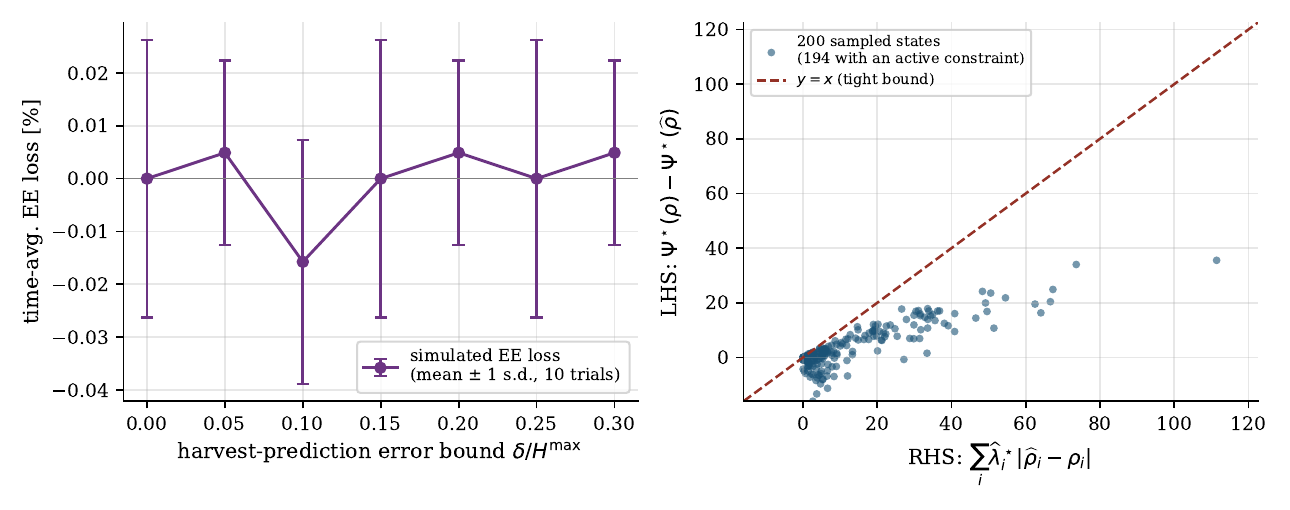}
\caption{Left: time-average EE loss versus harvest-prediction error bound
$\delta/H^{\max}$ under a battery-stressed scenario ($V=50$, harvested
power scaled to $70\%$ of nominal); error bars are $\pm1$ standard
deviation over 10 trials. Right: direct validation of
Theorem~\ref{thm:sensitivity} -- measured fixed-subproblem value variation
$\widehat\Psi^{(r)\star}(\bm\rho)-\widehat\Psi^{(r)\star}(\widehat{\bm\rho})$
versus the dual-based bound
$\sum_i\widehat\lambda_i^{(r)\star}|\widehat\rho_i-\rho_i|$ over 200 sampled
operating points; points on or below the dashed $y=x$ line satisfy the
bound.}
\label{fig:harvest_error}
\end{figure*}
\subsection{Results Discussion}
\revstart
Figure~\ref{fig:ee_vs_V}(a) shows effective EE increasing from
$2.89$ to $4.41$~Mbit/s/W between $V=1$ and $V=5000$.
Panel~(b) uses the largest tested $V$ as its empirical reference.
The scaled gap grows over part of the sweep and vanishes at the
endpoint by definition. This experiment characterizes sensitivity
to $V$ on the tested configuration. Estimating a theoretical decay
rate would require an independently evaluated comparison rule and
its drift contribution.
\revstop
Fig. \ref{fig:battery_vs_V} plots battery-target deviation versus $V$. The initial analysis used a single aggregate metric
($E[|Z_i(t)|]$) without stating which theoretical object it corresponds
to, and labeled it a ``deficit,'' which is misleading since $Z_i(t)$ is
signed and $|Z_i(t)|$ grows whether the battery is dangerously low or
merely well above target. Fig.~\ref{fig:battery_vs_V} instead reports all
three natural quantities. $E[Z_i(t)]$ is negative throughout the tested
range and becomes more negative with $V$ (from $-59$~kJ at $V=1$ to
$-166$~kJ at $V=5000$): on average the battery sits \emph{above} its
target $\Omega_i$ in this well-provisioned nominal scenario, increasingly
so at large $V$. $E[|Z_i(t)|]$ grows similarly, from $108$~kJ to
$175$~kJ. The deficit-only component $E[Z_i(t)^+]$ instead
\emph{decreases} with $V$, from $25$~kJ at $V=1$ to $4.5$~kJ at $V=5000$:
at large $V$ the controller spends aggressively enough that the battery
rarely sits below target at all, consistent with the RF-hardware-ceiling
saturation discussed in Result~8. The 5th-percentile state of charge stays
above $57\%$ across the entire tested range (Table~\ref{tab:params}
gives the scenario settings), i.e. even the low tail of the battery trajectory
remains far from the depth-of-discharge floor in this nominal,
well-provisioned scenario.

Fig.~\ref{fig:trajectory} shows a representative six-orbit trajectory: the
battery charges outside eclipse and discharges during it in a regular
sawtooth pattern, oscillating between $163$ and $269$~Wh. The simulated
trajectory remains within $[E^{\min},E^{\max}]=[60,300]$~Wh, consistently
with \rev{the scalar battery model and its clipping convention}.

Fig.~\ref{fig:harvest_error} (left) injects a bounded harvest-estimation
error into $\widehat H_i(t)$ (Assumption~\ref{as:causal}) while the true
battery update always uses the true $H_i(t)$, and the controller's
causality bound uses the corresponding robust margin
(Eqs.~\eqref{eq:causality}--\eqref{eq:overflow}). The measured EE is
\emph{identical to floating-point precision} across several tested
$\delta/H^{\max}$ values (e.g. $\delta/H^{\max}=0.05, 0.20, 0.30$ all give
$3.21948\ldots$). This is not a plotting artifact: because the robust
margin uses $[\widehat H_i(t)-\delta_{H,i}]^+$, once $\delta_{H,i}$ is
large enough that this pessimistic bound is already clipped to zero during
the eclipse-adjacent, harvest-scarce portions of this deliberately stressed
scenario, further increases in $\delta_{H,i}$ cannot tighten the causality
bound any further there, and the resulting trajectory -- and hence EE -- is
exactly, not just approximately, unchanged. The residual variation across
the remaining values is within Monte Carlo noise ($\le0.03\%$, comparable
to the $0.017$--$0.026\%$ trial-to-trial standard deviation). We read this
as a genuine (if scenario-specific) property of the robust, conservative
formulation rather than a numerical confirmation of the exact constant in
Theorem~\ref{thm:sensitivity} -- which Fig.~\ref{fig:harvest_error} (right)
addresses directly instead.


An EE-loss-versus-error plot alone does not test
Theorem~\ref{thm:sensitivity}'s specific claim, which concerns the value of
a fixed convex subproblem rather than the achieved EE. For the reduced-order
control model, the operating-point subproblem is convex and contains no
matrix-level SCA update; the experiment therefore evaluates
Theorem~\ref{thm:sensitivity} in this reduced convex setting. Over 200
sampled operating points, the resource constraint is active in 194 cases,
and the measured value variation satisfies the dual-based upper bound in
all samples within the numerical solver tolerance of $10^{-3}$.

\begin{table*}[t]
\centering
\caption{\rev{Six-policy comparison under nominal and battery-stressed scenarios (10 trials per cell). Battery minima are averaged across trials in the clipped scalar model.}}
\label{tab:baselines}
\small
\resizebox{\textwidth}{!}{%
\begin{tabular}{@{}llccccc@{}}
\toprule
Policy & Scenario & EE [Mbit/s/W] & Mean trial min. [Wh] & Near-depletion [\%] & Delivered [Gbit] & Shortfall [W] \\
\midrule
Lyapunov (proposed) & nominal  & 3.26 & 157.7 & 0.0  & 65072 & 0.00 \\
Lyapunov (proposed) & stressed & 2.27 & 92.7  & 0.0  & 22533 & 0.00 \\
Myopic               & nominal  & 4.41 & 160.1 & 0.0  & 62096 & 0.00 \\
Myopic               & stressed & 4.07 & 60.0  & 19.2 & 56312 & 0.00 \\
Fixed 20~W           & nominal  & 4.35 & 154.0 & 0.0  & 63691 & 0.00 \\
Fixed 20~W           & stressed & 4.04 & 60.0  & 20.1 & 56163 & 2.47 \\
Fixed 35~W           & nominal  & 4.03 & 109.0 & 0.0  & 78383 & 0.00 \\
Fixed 35~W           & stressed & 3.49 & 60.0  & 36.5 & 57591 & 9.44 \\
Fixed 50~W           & nominal  & 3.59 & 60.0  & 8.7  & 83965 & 2.59 \\
Fixed 50~W           & stressed & 3.03 & 60.0  & 48.2 & 55309 & 18.78 \\
Equal-power (53.6~W) & nominal  & 3.47 & 60.0  & 12.2 & 83309 & 4.24 \\
Equal-power (53.6~W) & stressed & 2.93 & 60.0  & 50.4 & 54637 & 21.23 \\
\bottomrule
\end{tabular}%
}
\end{table*}

\begin{figure}[t]
\centering
\includegraphics[width=0.85\linewidth]{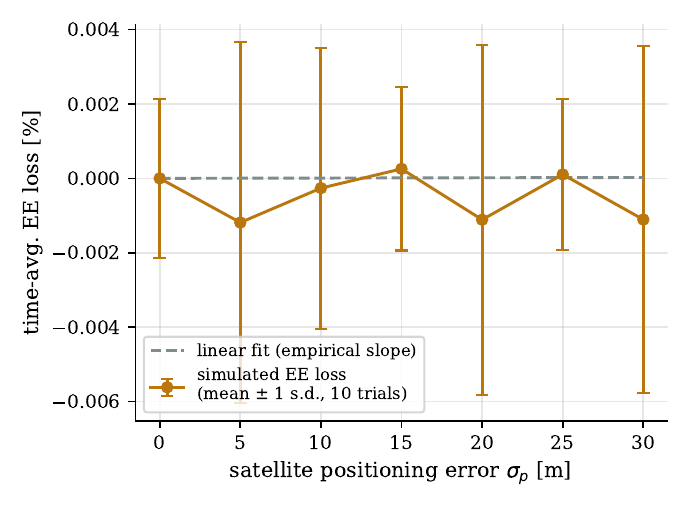}
\caption{Time-average EE loss versus satellite positioning error
$\sigma_p$, propagated through the actual 3-D satellite/UE geometry into
the large-scale fading coefficient $\alpha_{i,j}$ (Eq.~\eqref{eq:lsfc});
error bars are $\pm1$ standard deviation over 10 trials.}
\label{fig:pos_error}
\end{figure}
\revstart
Figure~\ref{fig:pos_error} applies position errors to the scalar
experiment through large-scale fading. In this experiment the
observed change in EE remains within trial-to-trial variation over
the tested range. The scalar model has no array phase or coherent
interference terms, so this sweep isolates the large-scale-fading
path and does not test the full position sensitivity in
Theorem~\ref{thm:possens}.
\revstop

\label{sec:r6}
\begin{figure*}[t]
\centering
\includegraphics[width=0.98\linewidth]{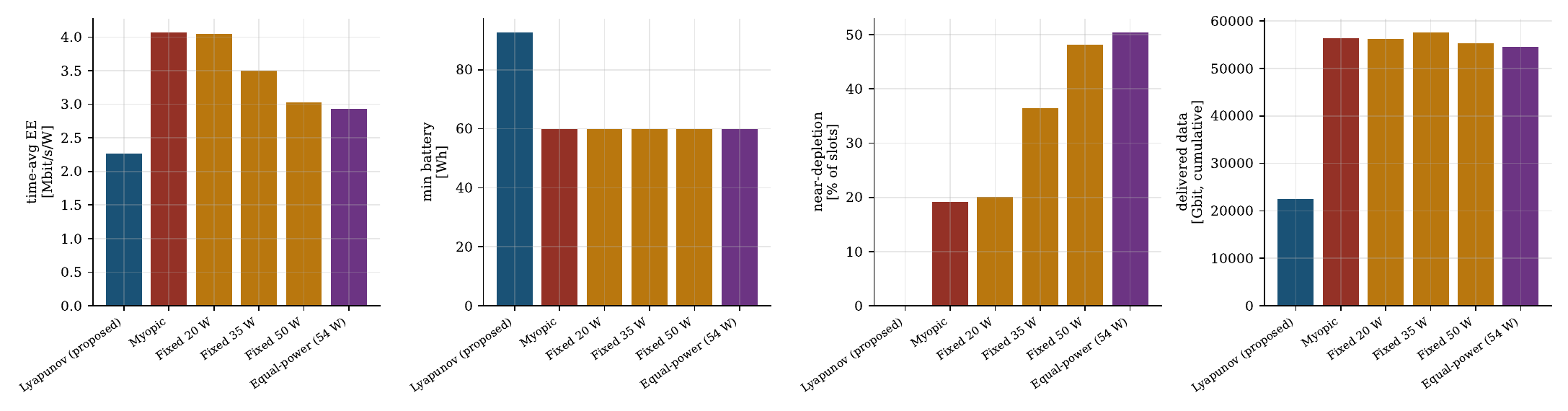}
\caption{Proposed Lyapunov-based controller vs.\ four baselines under a
battery-stressed scenario (harvested power scaled to $70\%$ of nominal,
$V=20$), averaged over 10 trials per policy.}
\label{fig:comparison}
\end{figure*}
Fig.~\ref{fig:comparison} compares the proposed framework with three representative
non-queue-aware control strategies: a myopic policy, fixed-power-budget
policies, and an equal-power policy. All baseline policies use the scalar simulator’s cap on controllable power
$e_i(t)\le e_i^{\max}(t)$ (Eq.~\eqref{eq:causality}) as every other policy
in this paper -- none of the baselines below is permitted to draw more
power than the battery can physically supply at that instant, even though
none of them is aware of the battery when \emph{deciding} how much to draw.
\begin{itemize}
\item \textbf{Myopic:} the same per-slot quadratic-transform solver, with
the battery-queue weight $Z_i(t)$ set to zero.
\item \textbf{Fixed power budget:} $e_i(t)=\min(\rho,e_i^{\max}(t))$ for a
design-time constant $\rho$, representative of the fixed-per-satellite-budget
assumption of prior cell-free NTN EE designs (Section~\ref{sec:related}).
We sweep $\rho\in\{20,35,50\}$~W to test sensitivity to this design
choice, rather than picking one value that might flatter the comparison.
\item \textbf{Equal power:} $e_i(t)=\min(\rho_{\mathrm{eq}},e_i^{\max}(t))$
with a single constant $\rho_{\mathrm{eq}}$ shared by every satellite
(equal split at the satellite level, not the per-UE level), calibrated as
the empirical average per-satellite causality bound over a representative
run of the stressed scenario under a neutral reference policy, giving
$\rho_{\mathrm{eq}}=53.6$~W.
\end{itemize}
Both new baseline families report an additional \emph{shortfall}
diagnostic, $\max(0,\rho-e_i^{\max}(t))$ (or $\rho_{\mathrm{eq}}$ in place
of $\rho$): how often the design's target exceeds what the battery can
physically deliver that instant, since neither baseline adapts to find
out. \emph{Near-depletion} is defined, as before, as
$E_i(t)\le(1+0.1)E_i^{\min}=66$~Wh, i.e. within $10\%$ of the
depth-of-discharge floor.

Table~\ref{tab:baselines} reports all six policies under both the nominal
and battery-stressed scenarios of Section~\ref{sec:setup}. Three findings
stand out. First, under the nominal scenario, a fixed budget set too close
to the $60$~W RF hardware ceiling is \emph{not safe even when
well-provisioned}: the $50$~W design and the $53.6$~W equal-power design
both show nonzero near-depletion ($8.7\%$ and $12.2\%$) purely because
neither adapts to the eclipse cycle, while every adaptive or
conservatively-fixed policy stays at $0\%$. Second, under the stressed
scenario shown in Fig.~\ref{fig:comparison}, every non-battery-aware
policy -- myopic \emph{and} all three fixed/equal-power variants -- reaches
the depth-of-discharge floor, while only the proposed Lyapunov controller maintains a higher observed reserve (\rev{mean trial minimum of $92.7$~Wh versus $60.0$~Wh for every other policy in the clipped model}); this is the central point the comparison is meant to make: it is
\rev{the queue term that improves the observed reserve in this scalar comparison}. Third,
the ranking among the non-battery-aware policies is not what a naive
mapping from ``more conservative cap'' to ``safer'' would predict: myopic
($19.2\%$ near-depletion, $4.07$~Mbit/s/W EE) weakly \emph{dominates} the
$20$~W fixed budget ($20.1\%$ near-depletion, $4.04$~Mbit/s/W EE) on both
metrics simultaneously, because myopic still re-solves an EE-optimal
allocation every slot (just without the battery-queue term) and so
partially, incidentally tracks favorable conditions, whereas a truly fixed
budget cannot. This must not be read as a general result that myopic
dominates fixed-budget designs -- it is a property of this particular
fixed value relative to this particular scenario -- but it does show that
``optimizing without battery awareness'' and ``capping power without
optimizing'' are meaningfully different failure modes, not interchangeable
stand-ins for the same thing.

We also flag an honest limitation of this comparison: at $V=20$ (chosen to
demonstrate the stressed-scenario reserve improvement), the proposed controller
has \emph{lower} EE than every other policy, including the crude
equal-power baseline, under the \emph{nominal} scenario ($3.26$~Mbit/s/W
vs.\ $3.47$--$4.41$~Mbit/s/W). This is consistent with, and explained by,
Result~8: $V=20$ sits well inside the queue-limited trade-off region, so
the controller is deliberately holding reserve that the nominal scenario
does not actually need. A real deployment would presumably select $V$
based on an assessed risk level (e.g.\ time to next eclipse, panel health),
which this fixed-$V$ comparison does not test, and which we leave to
future work.

\begin{figure}[t]
\centering
\includegraphics[width=0.85\linewidth]{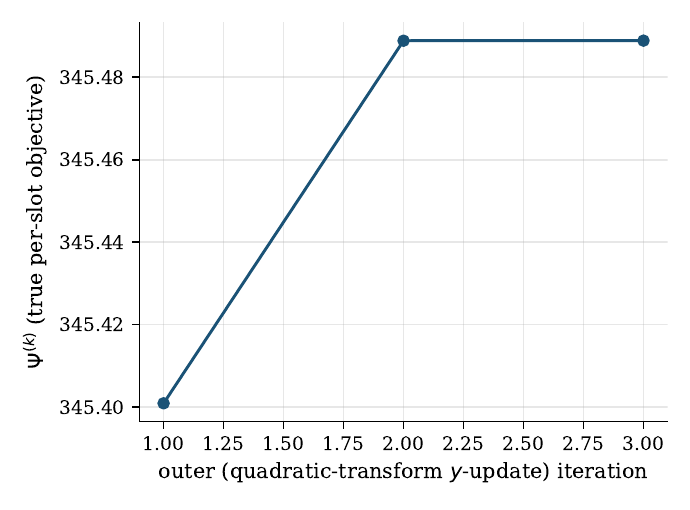}
\caption{Convergence of the true per-slot objective $\Psi^{(k)}$ under the
quadratic-transform outer ($y$-update) iteration, for one representative
slot ($V=100$, all batteries at $60\%$ of capacity).}
\label{fig:convergence}
\end{figure}
Fig.~\ref{fig:convergence} reports convergence of the true
queue-penalized objective rather than the EE ratio alone. This is necessary
because a ratio-only trace would not represent the additive
battery-weighted term in the per-slot objective. The objective is
nondecreasing at every step and converges within three iterations to the
prescribed relative tolerance of $10^{-6}$. This provides a reduced-order
check of the monotonic behavior established in
Proposition~\ref{prop:conv}, but does not constitute validation of the full
matrix-level beamforming, group-sparsity, and conic-optimization procedure.

\begin{figure}[t]
\centering
\includegraphics[width=0.85\linewidth]{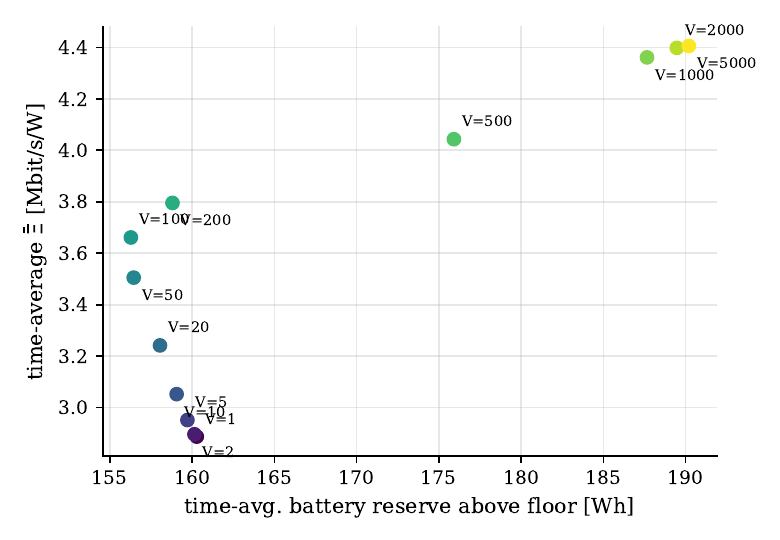}
\caption{Achieved operating points (time-average EE vs.\ time-average
battery reserve above the floor) as $V$ is swept from $1$ to $5000$.}
\label{fig:pareto}
\end{figure}
Fig. \ref{fig:pareto} discusses the EE--battery--reserve trade-off. It can be seen that the reserve first decreases as $V$ grows from $1$ to $\sim100$ ($160\to156$~Wh
above the floor) while EE rises -- a genuine trade-off region -- but then
increases again for $V\gtrsim200$, reaching $190$~Wh at $V=5000$. As
discussed for Result~2, this is because at large $V$ the controller's
chosen communication power saturates the $60$~W RF hardware ceiling rather
than being limited by the battery queue; once that ceiling is the binding
constraint, further increases in $V$ no longer draw down the battery any
faster. We do not present this as a clean, globally monotonic
\rev{scaling law. The main observed reserve-sensitive region is $V\in[1,100]$}
for this hardware configuration, and the large-$V$ regime is
hardware-limited saturation, a distinct (and separately worth noting)
phenomenon.

\vspace{0.5cm}
\noindent{\bf CRediT Authorship Contribution Statement}

Wali Ullah Khan: Writing – original draft, Methodology, Investigation, Conceptualization. Marwa Obayya: Writing – review \& editing, Supervision, Methodology, Conceptualization. Amal M Aqlan: Methodology, Conceptualization. Essa Alyounis: Writing – review \& editing. Randa Allafi: Methodology, Investigation, Conceptualization. Mutasim Al Sadig: Writing – review \& editing, Supervision, Methodology.

\vspace{0.5cm}
\noindent{\bf Declaration of Competing Interest} 

The authors declare that they have no known competing financial interests or personal relationships that could have appeared to influence the work reported in this paper.

\vspace{0.5cm}
\noindent{\bf Data Availability} 

Data will be made available on request.

\begin{figure*}[t]
\centering
\includegraphics[width=0.96\linewidth]{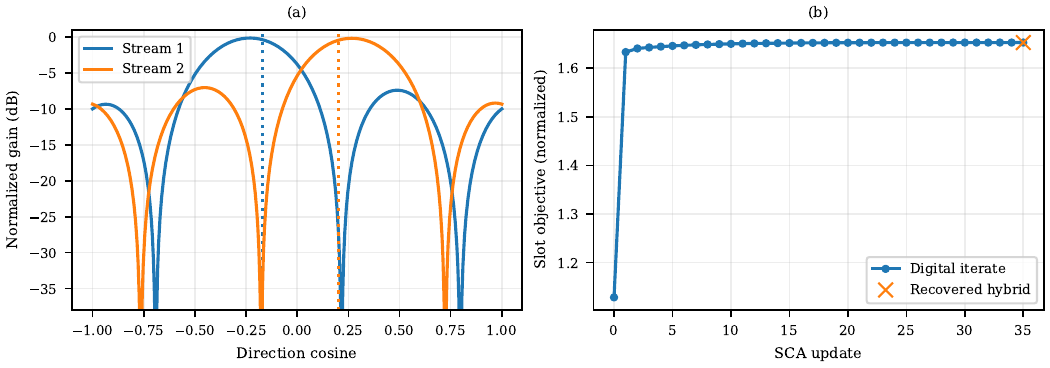}
\caption{\rev{Fixed-cluster matrix snapshot. (a) Beam cuts at satellite 1,
with dotted lines at the intended UE directions. (b) True slot objective
at each fixed-weight update and after hybrid recovery.}}
\label{fig:matrix}
\end{figure*}

\subsection{Fixed-Cluster Matrix-Level Snapshot}
\label{sec:matrix}
\revstart
We use two satellites, two single-antenna UEs, and $4\times4$
half-wavelength UPAs with two RF chains at each satellite. Both UEs are
served by both satellites. The horizontal direction cosines are
$(-0.17,0.20)$ at satellite 1 and $(-0.12,0.26)$ at satellite 2.
The experiment uses constant compensated LoS phases, independent
diffuse components, a $95\%$ LoS power fraction, and normalized link
powers $[1,0.75;0.8,1]$. Receiver noise is $0.025$, the per-satellite
PA-power cap is $2$, the total fixed communication power is $4$,
and the minimum spectral efficiency is $0.2$ bit/s/Hz per UE. The
normalized queue weights are $(0.06,-0.03)$ at $V=1$. We omit
forwarding in this snapshot and choose the energy bounds so that
the upper load limit reduces to the PA-power cap and the lower
load limit is inactive. All numerical powers are normalized.
The analog matrix consists of the two unit-norm constant-modulus
steering columns at each satellite. The digital iterate is solved
in their span, which contains the channel vectors in this example.
We initialize with half-amplitude steering beams and use SLSQP
for the fixed-weight convex minorization updates. Least-squares
recovery on the same analog matrix gives the recovered hybrid beams.

Figure~\ref{fig:matrix}(a) shows the two recovered beams at satellite 1
and marks the two UE directions. Network coherent interference
powers at UEs 1 and 2 decrease from $0.0867$ and $0.0832$ to
$8.21\times10^{-5}$ and $8.42\times10^{-5}$, respectively.
The recovered SINRs are $12.09$ and $12.00$~dB. Panel~(b) shows the
true slot objective increasing from $1.1282$ to $1.6530$ over 35
updates. The recovered objective agrees with the final digital value
to numerical precision because the iterates already lie in the
selected analog span. A separate bound on the slot optimum uses
the total normalized PA power $p\in[0,4]$. Cauchy--Schwarz with
receiver noise gives $\Gamma_j\le 1.71p_j/0.025$ for stream
power $p_j$, and the independent diffuse term gives
$\Gamma_j\le38$. Jensen's inequality therefore bounds the
sum rate by $2\log_2(1+\min\{34.2p,38\})$. The queue reward
is at most $0.03\min\{p,2\}$. Evaluating these increasing
numerators at the right endpoint of each $0.001$-wide interval
and the EE denominator at its left endpoint gives a slot
upper bound of $2.1020$ after rounding upward. The resulting
gap bound is $0.4490$ for this snapshot. Both recovered PA powers and both UE rates
satisfy the snapshot constraints. This test illustrates spatial
selectivity and monotonic improvement for the fixed-cluster static
case. It does not evaluate cluster selection, general hybrid projection
error, or time-varying phase and forwarding loads.
\revstop

\section{Conclusion}

\revstart
The battery-aware controller couples downlink scheduling with orbital
harvesting and onboard loads. Its robust constraints preserve the
physical battery interval whenever a feasible implemented action is
available, while its utility guarantee is conditional on the reference
drift and per-slot implementation gap. The long-horizon scalar study
quantifies the reserve--efficiency trade-off in the tested scenarios.
The fixed-cluster matrix snapshot illustrates beam directionality and
the objective trace in a static channel. It leaves open the performance
of joint cluster recovery and hybrid projection in time-varying orbital
channels.
\revstop

\end{document}